\documentclass[sigconf,nonacm]{acmart}
\AtBeginDocument{%
  }

\setcopyright{cc}
\setcctype{by}
\acmJournal{PACMMOD}
\acmYear{2026} \acmVolume{4} \acmNumber{6 (SIGMOD)} \acmArticle{340}
\acmMonth{12} \acmDOI{10.1145/3856335}

\usepackage[a-2b]{pdfx}   

\usepackage{amsmath}
\usepackage{siunitx}
\usepackage{booktabs, multirow}
\usepackage{mathtools}
\usepackage{graphicx} 
\usepackage{url}
\usepackage{enumitem} 

\usepackage{stmaryrd} 

\usepackage{upgreek} 
\usepackage{bbding} 

\usepackage{listings} 
\usepackage{tabularx} 
\usepackage{booktabs} 
\usepackage{multirow}
\usepackage{hhline} 
\usepackage{diagbox}
\usepackage{pgfplotstable} 

\usepackage{xcolor,colortbl}    

\usepackage{tikz} 
\usetikzlibrary{matrix}
\usetikzlibrary{calc}
\usetikzlibrary{math}
\usetikzlibrary{arrows.meta,positioning}
\usetikzlibrary{intersections, pgfplots.fillbetween}

\usepackage{easybmat} 

\usepackage{adjustbox}
\def\eox{\unskip\kern 10pt{\unitlength1pt\linethickness{.4pt}$\diamondsuit${}}} 

\usepackage{rotating}		

\usepackage{xspace} 

\usepackage{subcaption} 
\newcommand{\hide}[1]{}

\usepackage[ruled,noend,linesnumbered]{algorithm2e} 

\usepackage{setspace} 

\DontPrintSemicolon     
\SetNlSty{}{}{}                
\SetAlgoInsideSkip{smallskip}   

\SetAlFnt{\small}			
\SetAlCapFnt{\small}		
\SetAlCapNameFnt{\small}

\hypersetup{                                                            
	pdfpagemode=UseNone,               
    pdfstartview=Fit,
    pdfpagelayout=SinglePage,       
    bookmarks=false,
    bookmarksnumbered=true,
    bookmarksopen=false,             
    pdftex=true,
    unicode,
    colorlinks=true,       
    linkcolor=blue,          
    citecolor=cyan,  
    filecolor=magenta,      
     urlcolor=blue           
     plainpages=false,           
     hypertexnames=true,         
     pdfpagelabels=true,         
     hyperindex=true,
     naturalnames=false     
}

\usepackage[capitalise,nameinlink]{cleveref} 

\usepackage{aliascnt}  	

\newtheorem{example}{Example}

\newtheorem{definition}{Definition}
\newtheorem{lemma}{Lemma}

\newtheorem{observation}{Observation}

\AtEndPreamble{%
    \hypersetup{colorlinks,
      linkcolor=purple,
      citecolor=blue,
      urlcolor=ACMDarkBlue,
      filecolor=ACMDarkBlue}}

\usepackage{tcolorbox}		
\tcbuselibrary{breakable,skins}		

\tcbset{examplestyle/.style={
		enhanced jigsaw,	
		colback=blue!10,	
		colframe=blue!10,	
		arc=0mm,
		boxrule=0pt,		
		left=2mm,
		right=2mm,
		left skip= 0mm,  
		right skip= 0mm, 
		top=3pt,		
		bottom=3pt,		
		breakable,		
		parbox = false,		
		before={\par\pagebreak[0]\vspace{1mm}\parindent=0pt},		
		after={\par\pagebreak[0]\vspace{1mm}\parindent=0pt},				
		bottomrule = 0mm,
		boxsep = 0mm,					
		topsep at break=0pt,			
		bottomsep at break=0pt,			
		pad at break=0mm,
		pad before break=1mm,		
		pad after break=1mm,		
		bottomrule at break=0mm,
		toprule at break=0mm,		
		}}

\tcbset{querystyle/.style={
		enhanced jigsaw,	
		colback=blue!10,	
		colframe=blue!10,	
		arc=0mm,
		boxrule=0pt,		
		left=1mm,
		right=1mm,
		left skip= 0mm,  
		right skip= 0mm, 
		top=2pt,		
		bottom=2pt,		
		breakable,		
		parbox = false,		
		before={\par\pagebreak[0]\vspace{1mm}\parindent=0pt},		
		after={\par\pagebreak[0]\vspace{1mm}\parindent=0pt},				
		bottomrule = 0mm,
		boxsep = 0mm,					
		topsep at break=0pt,			
		bottomsep at break=0pt,			
		pad at break=0mm,
		pad before break=1mm,		
		pad after break=1mm,		
		bottomrule at break=0mm,
		toprule at break=0mm,		
		}}

\usepackage{footnote}

\tcolorboxenvironment{example}{examplestyle}
\tcolorboxenvironment{query}{querystyle}

\makeatletter
\DeclareRobustCommand*\uell{\mathpalette\@uell\relax}
\newcommand*\@uell[2]{
  \setbox0=\hbox{$#1\ell$}
  \setbox1=\hbox{\rotatebox{10}{$#1\ell$}}
  \dimen0=\wd0 \advance\dimen0 by -\wd1 \divide\dimen0 by 2
  \mathord{\lower 0.1ex \hbox{\kern\dimen0\unhbox1\kern\dimen0}}
}
\makeatother

\usepackage{marginnote}

\renewcommand{\epsilon}{\varepsilon} 

\renewcommand{\O}{{\mathcal{O}}} 

\usepackage{scalerel}
\usepackage{tikz}
\usetikzlibrary{svg.path}

\definecolor{orcidlogocol}{HTML}{A6CE39}
\tikzset{
  orcidlogo/.pic={
    \fill[orcidlogocol] svg{M256,128c0,70.7-57.3,128-128,128C57.3,256,0,198.7,0,128C0,57.3,57.3,0,128,0C198.7,0,256,57.3,256,128z};
    \fill[white] svg{M86.3,186.2H70.9V79.1h15.4v48.4V186.2z}
                 svg{M108.9,79.1h41.6c39.6,0,57,28.3,57,53.6c0,27.5-21.5,53.6-56.8,53.6h-41.8V79.1z M124.3,172.4h24.5c34.9,0,42.9-26.5,42.9-39.7c0-21.5-13.7-39.7-43.7-39.7h-23.7V172.4z}
                 svg{M88.7,56.8c0,5.5-4.5,10.1-10.1,10.1c-5.6,0-10.1-4.6-10.1-10.1c0-5.6,4.5-10.1,10.1-10.1C84.2,46.7,88.7,51.3,88.7,56.8z};
  }
}

\RequirePackage{etoolbox}
\DeclareRobustCommand\orcidicon[1]{\href{https://orcid.org/#1}{\mbox{\scalerel*{
\begin{tikzpicture}[yscale=-1,transform shape]
\pic{orcidlogo};
\end{tikzpicture}
}{|}}}}

\usetikzlibrary{automata,positioning,arrows.meta}
\usepackage{svg}

\usepackage{pgfplots}
\usepgfplotslibrary{groupplots}
\pgfplotsset{compat=1.18}

\newcommand{\ourabstraction}{\textsc{ReCAP}\xspace}

\newcommand{\init}{\texttt{init\_d}\xspace}
\newcommand{\update}{\texttt{update\_d}\xspace}
\newcommand{\finalize}{\texttt{finalize\_d}\xspace}
\newcommand{\isvalid}{\texttt{is\_viable\_d}\xspace}
\newcommand{\isvalidfinal}{\texttt{is\_viable\_d\_final}\xspace}
\newcommand{\merged}{\texttt{merge\_d}\xspace}
\newcommand{\ismergable}{\texttt{is\_mergable}\xspace}

\newcommand{\true}{\texttt{true}\xspace}
\newcommand{\false}{\texttt{false}\xspace}

\newcommand{\Bitcoin}{\textsc{Bitcoin}\xspace}

\usepackage{booktabs} 
\graphicspath{{./figs/}}
\usepackage{numprint}
\usepackage{multirow}
\usepackage{makecell}

\usepackage{balance}
\usepackage{tikz}
\usepackage{tikz,tikz-qtree,tikz-qtree-compat,tikz-3dplot}
\usetikzlibrary{shapes,decorations.markings,arrows.meta,fit}
\newcommand{\NEO}{\textsc{Neo4j}\xspace}
\newcommand{\KUZU}{\textsc{Kùzu}\xspace}
\newcommand{\MEMG}{\textsc{Memgraph}\xspace}
\newcommand{\DUCKDB}{\textsc{DuckDB}\xspace}
\newcommand{\SYSX}{\textsc{System X}\xspace}
\newcommand{\FinBench}{\textsc{FinBench}\xspace}

\usepackage{xcolor}
\definecolor{colIP}    {HTML}{E74C3C}
\definecolor{colOUT}   {HTML}{27AE60}
\definecolor{colDuckDB}{HTML}{E67E22}
\definecolor{colNeo4j} {HTML}{8E44AD}
\definecolor{colMemgraph}{HTML}{935116} 
\definecolor{colKuzu}  {HTML}{2980B9}
\definecolor{colReCAP} {HTML}{16A085}
\definecolor{colSysX} {HTML}{FF0000}

\definecolor{dkgreen}{rgb}{0,0.6,0}

\makeatletter
\newcommand{\markZwicky}[1][]{\pgfutil@ifnextchar({\mark@Zwicky{#1}}{\mark@Zwicky{#1}()}}
\def\mark@Zwicky#1(#2)#3{%
   \tikz[every Zwicky picture,#1]{%
     \node[every Zwicky node,draw=none,inner sep=+\z@,outer sep=+\z@] {#3};
     \def\tikz@Mark@name{#2}%
     \ifx\tikz@Mark@name\pgfutil@empty\else
       \tikzset{every Zwicky node/.append style={name={#2}}}%
     \fi
     \node[every Zwicky node,overlay] {\phantom{#3}};
   }%
}
\newcommand{\tikzZwicky}[1][]{%
  \def\tikz@Zwicky@args{#1}%
  \let\tikz@Zwicky@list\pgfutil@gobble
  \let\tikz@Zwicky@first\pgfutil@empty
  \pgfutil@ifnextchar(\tikz@Zwicky@collect\tikz@Zwicky@finish
}
\def\tikz@Zwicky@collect(#1){%
  \ifx\tikz@Zwicky@first\pgfutil@empty
    \edef\tikz@Zwicky@first{#1}%
  \else
    \edef\tikz@Zwicky@list{\tikz@Zwicky@list,#1}%
  \fi
  \pgfutil@ifnextchar(\tikz@Zwicky@collect\tikz@Zwicky@finish
}
\def\tikz@Zwicky@finish{%
  \tikz[remember picture,overlay]
    \draw[every Zwicky connector,/expanded=\tikz@Zwicky@args]
      (\tikz@Zwicky@first) [/expanded={@Zwicky@list/.list={\tikz@Zwicky@list}}] [every Zwicky connect finish/.try];
}
\pgfkeys{/expanded/.code={\edef\pgfkeys@temp{{#1}}\expandafter\pgfkeysalso\pgfkeys@temp}}
\makeatother
\tikzset{
  @Zwicky@list/.style={insert path={to[every Zwicky connector how/.try] (#1)}},
  every Zwicky picture/.style={
    baseline,
    remember picture,
  },
  every Zwicky node/.style={
    remember picture,
    anchor=base,
    inner sep=+2pt
  },
  every Zwicky connector/.style={
    ultra thick,
    red!80!black,
    draw opacity=.5,
    line cap=round,
    line join=round
  }
}

\makeatletter
\let\oldnl\nl
\newcommand{\nonl}{\renewcommand{\nl}{\let\nl\oldnl}}
\makeatother

\makeatletter
\newcommand{\algocf@VslineMine}[1]{
  \par\nointerlineskip
  \algocf@push{\skiprule}
  \hbox{\vrule
    \vtop{\algocf@push{\skiptext}
      \vtop{\algocf@addskiptotal\advance\hsize by -\skiplength #1}}}
  \algocf@pop{\skiprule}}
\newcommand\Block[1]{%
    \algocf@VslineMine{#1}%
}
\makeatother

\begin{document}

\title{Efficient Path Query Processing in Relational Database Systems}

\author{Diego Rivera Correa}
\affiliation{%
  \institution{Northeastern University}
  \city{Boston}
  \country{USA}}
\email{rivera.di@northeastern.edu}

\author{Mirek Riedewald}
\affiliation{%
  \institution{Northeastern University}
  \city{Boston}
  \country{USA}}
\email{m.riedewald@northeastern.edu}

\begin{abstract}

Path queries are crucial for property graphs, and there is growing interest in supporting not only regular expressions over labels, but also constraints on the property values of vertices and edges. Efficient evaluation of such queries requires early filtering of intermediate results, but state-of-the-art (SOTA) graph and relational DBMS struggle to effectively exploit property constraints. We introduce the notion of negatively stable constraints and propose \ourabstraction, a novel abstraction that leverages negative stability for holistically specifying early filtering opportunities in path queries. To create a \ourabstraction it suffices to (1) specify the regular expression for the label constraint (as in the MATCH clause of Cypher or GQL), (2) define functions to manage state (like for a user-defined aggregate), and (3) define a predicate to identify early filtering opportunities on the state. We also design a default \ourabstraction for query conditions that are not amenable to early filtering---in which case our technique behaves like SOTA graph and relational DBMS. \ourabstraction semantics are defined as a recursive SQL query template that can be instantiated and executed in a standard relational DBMS. When early filtering opportunities on property constraints exist, our implementation using \DUCKDB achieves speedups up to 230,000$\times$ over the competitors.

\end{abstract}






\setlength{\textfloatsep}{6pt plus 2pt minus 2pt}   
\setlength{\intextsep}{6pt plus 2pt minus 2pt}      
\setlength{\floatsep}{6pt plus 2pt minus 2pt}       
\setlength{\abovecaptionskip}{3pt}                  
\setlength{\belowcaptionskip}{2pt}

\newpage

\setcounter{page}{1}

\maketitle

\section{Introduction} \label{sec:intro}

Many problems in diverse domains can be modeled as graphs, where vertices
represent entities and edges the relationships between them.
Examples include social, biological, and financial transaction
networks. Analyzing such data often involves
querying for paths that satisfy constraints
\cite{angles2017foundations} including: regular expressions over edge labels,
monotonicity constraints on timestamps (e.g., to ensure causality),
aggregate constraints on weights (e.g., to ensure quality)
or combinations thereof. While path queries have been studied for decades,
producing many theoretical insights \cite{libkin2025querying},
integrating those techniques into practical data management systems is
``still in its experimental stage'' \cite{MhedhbiA24ModernTechniques}.
Recent graph query language standards like GQL \cite{francis2023researcher}
and SQL/PGQ \cite{DuckPGQ_SQL_PQL} enable users to design powerful
declarative path queries. Hence there is an urgent need for techniques
to efficiently execute such queries.

We focus on path-query processing in \emph{relational DBMS} for two reasons.
First, they are ubiquitous in practice. If a system that is already
up and running can efficiently support the target queries, this avoids the
additional complexity and cost of managing specialized engines for
each type of problem (and moving data between them).
Second, as Pavlo and Stonebraker \cite{stonebraker_what_2024} point out,
relational and graph DBMS will likely converge to a single engine
that supports both relational and graph data management.
That should certainly apply to path queries, which are
equivalent to join-heavy relational queries with selections and
projections \cite{jin2021making}. Hence our results also transfer
in principle to graph DBMS.

\begin{figure*}[tbp]
\centering
\includegraphics[width=\linewidth]{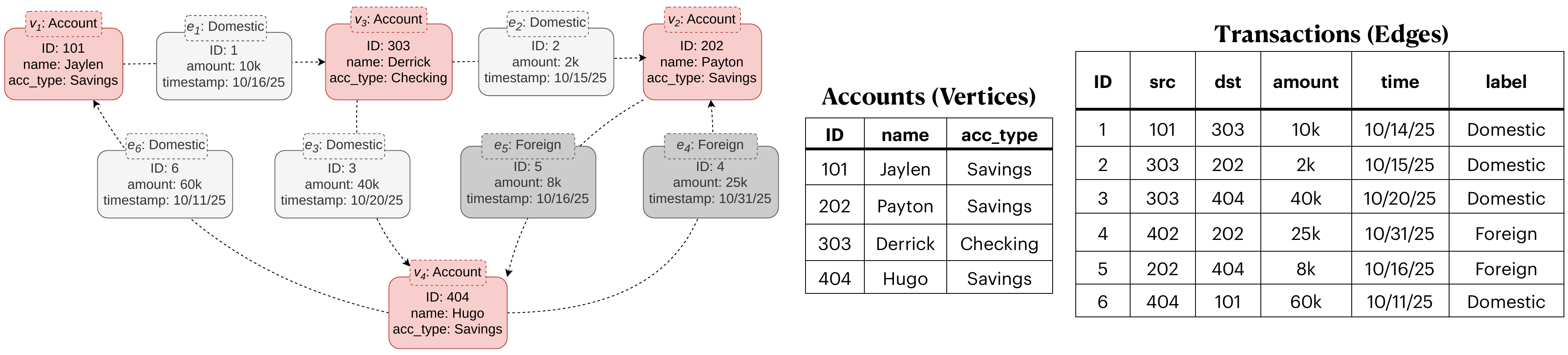}
\caption{On the left: Graph representing accounts (vertices) and transactions between them (edges).
The label is shown next to the ID; properties are shown inside the box.
On the right: Tabular representation of the graph.}
\label{fig:BitcoinGDB}
\end{figure*}

\begin{example}[Query $Q_A$]\label{ex:q1_bitcoin}
Consider a property graph of financial transactions, where vertices represent
accounts and edges the transactions between them (\Cref{fig:BitcoinGDB}).
A user wants to find all paths up to length $\ell$ starting from account $s$,
such that
(1) transactions along the path are in chronological order
(strictly increasing timestamps),
(2) no edge repeats (trail semantics), and
(3) the transferred amounts are relatively ``homogeneous'',
meaning that the difference between maximum and minimum transaction
amount along the path is upper bounded by some parameter $U$.
\end{example}

\begin{figure}[tbp]
    \centering
    \begin{tikzpicture}
    \begin{semilogyaxis}[
        xlabel        = {$\ell$},
        ylabel        = {Runtime (ms)},
        grid          = both,
        grid style    = {line width=0.3pt, draw=gray!30},
        major grid style = {line width=0.4pt, draw=gray!50},
        width         = 0.92\columnwidth,
        height        = 4.8cm,                              
        xmin=1.8, xmax=10.2,
        ymin=10,   ymax=2e7,
        xtick         = {2,3,4,5,6,7,8,9,10},
        ytick         = {1e1,1e2,1e3,1e4,1e5,1e6,1e7},
        yticklabels   = {$10^{1}$,$10^{2}$,$10^{3}$,$10^{4}$,
                         $10^{5}$,$10^{6}$,$10^{7}$},
        tick label style  = {font=\small},
        label style       = {font=\small},
        legend style      = {
            at={(0.5,1.02)}, anchor=south,        
            draw=gray!50,
            font=\small,
            legend columns=4,
            column sep=4pt,
            /tikz/every even column/.append style={column sep=8pt},
        },
        every axis plot/.append style = {mark size=3pt, line width=1pt},
    ]
    \addplot[color=colDuckDB, solid, mark=square]
        coordinates { (2,19) (3,116) (4,5555) (5,252294) };
    \addlegendentry{\DUCKDB}
    \addplot[color=colNeo4j,  solid, mark=diamond]
        coordinates { (2,363) (3,772) (4,23312) (5,1380168) };
    \addlegendentry{\NEO}
    \addplot[color=colKuzu,   solid, mark=triangle]
        coordinates { (2,142) (3,2549) (4,77263) (5,4228287) };
    \addlegendentry{\KUZU}
    \addplot[color=colReCAP,  solid, mark=o]
        coordinates {
            (2,15) (3,25) (4,78) (5,320)
            (6,1066) (7,2383) (8,4030) (9,5203) (10,5739)
        };
    \addlegendentry{\ourabstraction}
    \draw[<->, thick, black]
        (axis cs:5,800) -- (axis cs:5,100000)
        node[midway, right] {$\mathbf{788}\times$};
    \end{semilogyaxis}
    \end{tikzpicture}
    \caption{Runtime of \ourabstraction vs.\ SOTA competitors for
             varying path-length limits. All competitors scale poorly
             and exceed a 2-hour timeout for $\ell>5$.}
    \label{fig:scaling_runtimes}
\end{figure}

Executing such queries efficiently is far from trivial
as \Cref{fig:scaling_runtimes} indicates. We report running times of $Q_A$
on a \Bitcoin \cite{BitcoinOTC} transaction graph with 6k vertices and 36k edges.
Despite the small input, SOTA relational (\DUCKDB \cite{duckdb}) and
graph DBMS (\NEO \cite{Neo4j}, \KUZU \cite{feng2023kuzu}) fail to
scale with path length.\footnote{Experiments
with other SOTA DBMS yielded similar results and are omitted for brevity.}
A look at the query plans reveals that they suffer from huge intermediate
results because they do not push the conditions down effectively.
In contrast, our \ourabstraction abstraction (\Cref{sec:recap_definition})
enabled \DUCKDB to scale to more than double the path length.

\begin{figure}[tbp]
    \centering
    \begin{lstlisting}[language=SQL, escapechar=\%,
          basicstyle=\ttfamily\small,
          label={lst:sql_query_q1}, caption={SQL query for \Cref{ex:q1_bitcoin}}]
      %{\textbf{WITH}}% RECURSIVE Paths AS (
          SELECT s AS v,
                 0 AS path_length,
                 [] AS amounts,
                 [] AS times,
                 [] AS edge_ids
          UNION ALL 
          SELECT E.dst AS v,
                 P.path_length + 1,
                 P.amounts + [E.amount] AS amounts,
                 P.times + [E.time] AS times,
                 P.edge_ids + [E.id] AS edge_ids
          FROM Paths P
          JOIN Edges E ON P.v = E.src
          WHERE p.path_length < $\ell$
      )
      SELECT *
      FROM Paths 
      WHERE list_max(amounts) - list_min(amounts) <= U
      AND len(edge_ids) = len(list_distinct(edge_ids))
      AND NOT list_contains([times[i] <= times[i-1] 
          FOR i IN range(2, len(times)+1)], true)
    \end{lstlisting}
\end{figure}

To understand why the optimizers of \DUCKDB, \NEO, and \KUZU failed to push the
conditions down effectively, let us take a closer look at \DUCKDB.
(The situation is analogous for the graph DBMS.)
\Cref{lst:sql_query_q1} shows $Q_A$ in SQL. It uses a
recursive Common Table Expression (CTE) to define all paths up to length $\ell$,
accumulating edge IDs, amounts and timestamps along the path in arrays.
The outer WHERE clause filters paths based on the max-min constraint,
trail semantics, and the timestamp ordering condition.
Cypher-based graph DBMS would similarly rely on list processing over
the paths identified in the MATCH clause.

Consider the timestamp monotonicity condition expressed as
\texttt{NOT list\_contains(\ldots)} in \Cref{lst:sql_query_q1}.
It iterates through the values of
\emph{all} edges along the path and thus cannot be used for early
filtering on a path segment. Of course, some (not very deep) reasoning
can yield an effective early filtering strategy:
(1) The list constraint implies $e_i.\mathtt{time} < e_{i+1}.\mathtt{time}$
for each pair of adjacent path edges.
(2) When assembling the path left-to-right, maintain the last edge's
time value in a new column $\mathtt{last\_time}$ in Paths and
add condition $\mathtt{last\_time} < e.\mathtt{time}$
to the WHERE clause of the recursive member. Current optimizers
lack the analytical capabilities and rules to apply such
reasoning-based rewrites. Based on our analysis, this applies to
list-based constraints over recursive queries in general.

Recent work attempts to incorporate LLM reasoning functionality to identify more powerful rewrites
\cite{quite, genrewrite, lithe}.
While a human or LLM can produce a rewritten SQL query that
enables early filtering like for time monotonicity above, one still
must \emph{prove} that original and rewritten query are equivalent.
This is generally undecidable for recursive queries.
To address this problem in a principled manner, we propose an
abstraction called \ourabstraction
(\underline{\textbf{Re}}gular expressions, \underline{\textbf{C}}onstraints, and
\underline{\textbf{A}}ggregation \underline{\textbf{P}}atterns).
It formalizes the essence of which list-based property constraints can
be used for early filtering and how to exploit them effectively.
With \ourabstraction the rewritten query has a simpler
template-based structure where query equivalence reduces to
proving---typically via induction---invariants about paths
in a simple NFA-style automaton. In fact, without \ourabstraction,
one arguably would need to invent a representation like it to
convincingly prove equivalence between 2 path queries.

As a second major benefit, \ourabstraction naturally enables more
advanced execution strategies for path queries.
Consider finding paths starting at some vertex $s$ and ending at
another vertex $t$, such that the edges satisfy timestamp monotonicity
and their labels satisfy regex \texttt{A+ B+ C+}.
The above early-filtering construction for timestamp monotonicity
that maintains $\mathtt{last\_time}$ and checks condition
$\mathtt{last\_time} < e.\mathtt{time}$ in the WHERE clause
of the recursive member only works correctly for left-deep
plans that grow the path left-to-right from $s$.
What if label \texttt{B} occurs rarely and a more advanced
optimizer wants to use a plan
that finds matches for label pattern \texttt{B+} first,
then processes \texttt{A+} by searching forward from $s$ and
backward from the \texttt{B+} matches; and similarly forward
from the \texttt{B+} matches and backward from $t$ for
\texttt{C+}? For this to work, the \texttt{B+} matches must
keep both the first and the last edge's timestamp and when
connecting separate segments, the last time of the left
segment must be compared to the first time of the right.
In general, it is not clear how a user or LLM would tell
the optimizer how to combine path segments when the path
is not matched left-to-right.
With \ourabstraction, this is easy, requiring nothing more than
the definition of a straightforward function to merge state.

Third, like aggregates in relational algebra and SQL, common
\ourabstraction instances we refer to as
\emph{factorized} \ourabstraction enable greatly simplified
creation and maintenance of reusable libraries of
\emph{selective aggregates}. This way effort for development and
correctness proofs is amortized across all queries using the same
type of filtering pattern. In summary, we make the following major
contributions:

\textbf{(1)}
We identify \emph{negative stability} as a crucial structural property: if a constraint does not hold on a path segment, then no extension
of that segment can satisfy it. Our novel \ourabstraction abstraction
exposes negative stability so that a standard relational DBMS
can discard non-viable path segments the moment they are detected,
rather than extending them to full length and discarding them in the end.
\ourabstraction is fully general: it supports all combinations of
regular expressions over labels and data constraints on properties,
including aggregate constraints and regex-dependent thresholds.
When constraints do not admit early filtering, \ourabstraction reduces
to the query plans used by current SOTA systems that apply
list-based property constraints at the end.

\textbf{(2)}
Implementing a \ourabstraction is similar to specifying the MATCH-WHERE
block of a Cypher, GQL, or SQL/PGQ query, plus a user-defined
aggregate (UDA) on property values, with 7 functions to initialize,
maintain, and finalize state, and to determine if a path can be filtered early.
Through a \texttt{CASE} statement, these functions can modify their
behavior based on the state of the label pattern, thus enabling early
filtering based on complex constraints that involve dependencies
between the label pattern and property values.
When all cases collapse, i.e., a factorized \ourabstraction instance,
the aggregate-style functionality can be managed like UDAs in relational
DBMS, greatly simplifying their creation and correctness proofs.
We refer to these as \emph{selective aggregates}.

\textbf{(3)}
We demonstrate that \ourabstraction can be implemented
with little coding effort in SQL, using a standard recursive
select-project-join (SPJ) query.
We also propose effective optimizations including automaton-style processing
converted to efficient joins, dictionary flattening, and function inlining
that leverage the full power of SQL. Our \ourabstraction implementation
using \DUCKDB achieves orders of magnitude speedup and scales to significantly
longer paths compared to SOTA graph and relational DBMS.

\section{Problem Definition}
\label{sec:problem}

We study path queries over property graphs. For simplicity, we only discuss queries with conditions on edges;
extending our approach to vertices is straightforward.

\begin{definition}[Property Graph \cite{angles2017foundations}]
\label{def:property_graph}
A \emph{property graph} $G$ is a 5-tuple $(V, E, \rho, \lambda, \sigma)$ where
$V$ is a finite set of vertices,
$E$ is a finite set of edges with $V \cap E = \emptyset$,
$\rho: E \rightarrow (V \times V)$ is a total function that maps an edge to its
source and target vertex,
$\lambda: (V \cup E) \rightarrow L$ is a total function that assigns a label from
label set $L$ to each vertex and edge, and
$\sigma: (V \cup E) \times \mathrm{Prop} \rightarrow \mathrm{Val}$ is a
partial function that assigns properties to vertices and edges. Each property
is a key-value pair with a key from $\mathrm{Prop}$ and a value from $\mathrm{Val}$.
In addition to scalars, we also allow complex values
such as lists and dictionaries.
\end{definition}

\begin{example}
In \Cref{fig:BitcoinGDB}, $\rho(e_2) = (v_3, v_2)$, $\lambda(e_{4}) = \texttt{Foreign}$,
$\lambda(n_1) = \texttt{Account}$, and $\sigma(e_5, \texttt{amount}) = \texttt{8k}$.
\end{example}

\begin{definition}[path, segment, extension, word]
\label{def:path}
A \emph{path} $p$ of length $k$ is a sequence of vertices and edges
$v_0 e_1 v_1 \cdots e_k v_k$ for $k \ge 1$,
where $e_i \in E$ and $\rho(e_i) = (v_{i-1}, v_i)$ for each $1 \le i \le k$.
Any $p' = v_i e_{i+1} v_{i+1} \cdots e_j v_j, 0 \le i < j \le k$
is a \emph{segment} of $p$; and $p$ is an \emph{extension} of $p'$.
We write $p' \preceq p$ to denote that $p'$ is a segment of $p$.
We write $\Pi(G)$ for the set of all paths in $G$. $\Pi_\ell(G)$ denotes
the set of all paths in $G$ of length at most $\ell$.
The word $w(p)$ of a path $p$ is the concatenation of the labels of
each edge in $p$. 
\end{definition}

\begin{example}
In \Cref{fig:BitcoinGDB} the path  $p = v_3e_{2}v_2e_{5}v_4$ from Derrick's account to
Hugo's account, generates the word
$\texttt{Domestic} \cdot\texttt{Foreign}$.
\end{example}

\begin{definition}[regular expression, language]
\label{def:regex}
A \emph{regular expression} $R$ over a label set $L$ is defined
in the standard way~\cite{sipser1996introduction}.
The \emph{language recognized by $R$}, written $\mathcal{L}(R)$,
is the set of strings over $L$ generated by $R$. 
\end{definition}

\begin{definition}[path query]
\label{def:path_query}
A \emph{path query} $Q = (S, R, \varphi)$ over a property graph $G$
is defined by a set of start vertices $S \subseteq V$,
a regex $R$ over label set $L$,
and a predicate $\varphi$ defined for all paths in $G$.
It returns a set of paths in $G$ such that each result path
$p = v_0 e_1 v_1 \cdots e_k v_k$ satisfies the following properties:
\begin{itemize}[noitemsep,nolistsep]
  \item $v_0 \in S$,
  \item $w(p) \in \mathcal{L}(R)$, and
  \item $\varphi(p) = \true$.
\end{itemize}
A \emph{length-limited path query} $Q_\ell$ further requires $k \le \ell$.
\end{definition}

We support other variations of the path query: no start vertex given,
start vertex and end vertex given, or that the path must cross through
some specific vertices ``in the middle''. However, we will focus on path
queries that originate from a given start vertex $s$.
Our technique relies on classic finite automata (NFA):

\begin{definition}[NFA]
\label{def:NFA}
The NFA for path query $Q = (S, R, \varphi)$ over property graph $G$
is a 5-tuple $N = (\mathfrak{Q}, L, \delta, q_0, \mathfrak{Q}_F)$, consisting of
\begin{itemize}[noitemsep,nolistsep]
    \item A set of states $\mathfrak{Q}$,
    a single initial state $q_0 \in \mathfrak{Q}$, and
    a set $\mathfrak{Q}_F \subseteq \mathfrak{Q}$ of accepting states.
    \item A label set $L$ (the same as the property graph's label set).
    \item A transition function $\delta: \mathfrak{Q} \times L \rightarrow \mathcal{P}(\mathfrak{Q})$,
    where $\mathcal{P}(\mathfrak{Q})$ denotes the power set of $\mathfrak{Q}$.
\end{itemize}
The language recognized by $N$ is equivalent to the language specified by $R$,
i.e., $\mathcal{L}(N) = \mathcal{L}(R)$.
\end{definition}

Note that \Cref{def:path_query} includes cases that are
efficiently processed by SOTA graph DBMS, such as path queries
with (selective) label regex\footnote{Leading graph DBMS
\emph{do not support all regular expressions}, but our technique does.}
only, or those where property constraints can be attached to
individual edges within the regex pattern. For example, Cypher's
\texttt{MATCH (s) -[e1]-> (v1) -[e2 WHERE e1.time < e2.time]-> (t)}
enforces timestamp monotonicity between 2 named edges
of a 2-hop path. We focus on ``difficult'' instances with
iterating patterns where path length is not pre-determined.
Query $Q_A$'s regex pattern corresponds to
\texttt{MATCH ()-[]->\{1,\}()}, therefore one cannot express
timestamp monotonicity on its edges there~\cite{gheerbrant2024gql},
which forces Cypher implementations to resort to
list processing~\cite{gheerbrant2025dangers}.
We will refer to such constraint as \emph{list-based}.
To enable efficient processing of queries with list-based constraints,
one must identify structural properties that enable early filtering
of intermediate results that cannot lead to a query answer:

\begin{definition}[(Non-) viable path segments]\label{def:doomed}
  Let $Q = (S, R, \varphi)$ be a path query on property graph $G$.
  Path segment $p' \in \Pi(G)$ is \emph{non-viable} if none of its
  possible extensions can be a query answer;
  otherwise it is \emph{viable}.
\end{definition}

Whether a segment is viable depends on both label constraint
$R$ and property constraint $\varphi$. The former is determined
by the NFA: if the current state has no path to an accepting
state, the intermediate result is non-viable. For the latter,
the analogue is a constraint guaranteeing that once it fails
on a segment $p'$, it must fail on every extension of $p'$:

\begin{definition}[Negatively stable constraint]
\label{def:neg-stable}
  Let $\varphi$ be a predicate as defined in \Cref{def:path_query}.
  We say $\varphi$ is \emph{negatively stable} if for every
  path $p' \in \Pi(G)$ and every extension $p$ of $p'$:
  $\varphi(p') = \false \;\;\Rightarrow\;\; \varphi(p) = \false$.
\end{definition}

\begin{example}[Negatively Stable Constraints in $Q_A$]
\label{ex:Q_A_negatively_stable}
  All property constraints from $Q_A$ are negatively stable
  (monotonicity of timestamps, trail semantics, upper bound
  on aggregate range).
\end{example}

Clearly, if $\varphi(p') = \false$ for a negatively stable constraint
$\varphi$, then path segment $p'$ is non-viable and can be filtered out.
Fortunately, even if $\varphi$ is not negatively stable, it may ``contain''
a negatively stable constraint that can be leveraged for early filtering:

\begin{lemma}\label{lem:conj_stable}
    If $\varphi = \varphi_1 \land \varphi_2$, where $\varphi_1$ is
    negatively stable, then a path segment $p'$ is non-viable
    if $\varphi_1(p') = \false$.
\end{lemma}

\begin{example}
\label{ex:conj_stable}
Consider paths of even length with increasing timestamps.
This constraint is not negatively stable,
because a 3-edge path with timestamps $[1,2,3]$ violates it
(odd length), but its extension by appending an edge with
timestamp 4 satisfies it. Setting $\varphi_1$ to timestamp
monotonicity and applying \Cref{lem:conj_stable}, we can still
leverage timestamp monotonicity for early filtering, while
even path length can only be enforced on the full path.
\end{example}

\section{Running Example}
\label{sec:running_example}

We introduce an example query that highlights all features of \ourabstraction
and motivates the general abstraction.
\begin{example}[query $Q_B$]
\label{ex:full_recap}
In the financial-transaction graph from \Cref{ex:q1_bitcoin}, find all paths
from a given \texttt{Account} $s$, such that:
\begin{enumerate}[noitemsep,nolistsep]
  \item The labels match pattern $\texttt{Domestic}^+\texttt{Foreign}$, i.e., at least one
  \texttt{Domestic} followed by exactly one \texttt{Foreign} transaction.
  
  \item Consecutive \texttt{Domestic} transactions are executed within 2 days
  of each other, i.e., their timestamp difference falls in range $[-2,2]$ days.
  
  \item The last \texttt{Domestic} and the \texttt{Foreign} transaction are
  executed within 3 days of each other, i.e., their timestamp difference falls in range $[-3,3]$ days.
  
  \item No edge is repeated (trail semantics).
\end{enumerate}
For each result path, return the sequence of its edge IDs. 
\end{example}

Query $Q_B$ generalizes $Q_A$ by introducing a label regex, creating
a dependency between label pattern and property constraint (different timestamp
intervals), and generalizing the timestamp monotonicity. The latter allows
searching for suspicious transfers that attempt to
hide the money flow by inverting transaction order: To hide that \$10 move
from $A$ to $B$ to $C$, $B$ could first send \$10 to $C$, then soon after
get ``reimbursed'' by $A$.

It is easy to identify opportunities for early filtering when considering
the next transaction edge to be appended to a path:
\begin{itemize}[noitemsep,nolistsep]
\item Use an NFA to efficiently check the label pattern.

\item Maintain $\mathtt{last\_time}$, the time of the path's last transaction.
Filter out transaction $e$ if $e.\mathtt{time} - \mathtt{last\_time}$
falls outside the time range (2 or 3 days, respectively).

\item Maintain $\mathtt{edge\_ids}$, a list of IDs of all edges along the path.
Filter out transaction $e$ if $e.\mathtt{id} \in \mathtt{edge\_ids}$.
\end{itemize}
The challenges here are:
\begin{itemize}[noitemsep,nolistsep]
\item Prove correctness of the modifications.

\item How to efficiently implement an NFA in a relational engine?

\item How to implement early filters that depend on both the state of the label
pattern and the data values (e.g., time range 2 vs 3)? In the example,
it suffices to check $e.\mathtt{label}$, but in general it depends
on the state transition of the NFA, which the relational engine must keep
track of.

\item How to implement state that grows with path length,
like the list of edge IDs? In a recursive SQL query the schema is fixed, hence
state must fit into a fixed set of columns.

\item 
How to enable support for advanced query plans beyond left-deep
(left-to-right) path expansion, especially matching/combining path segments
and bi-directional search?

\end{itemize}

\section{The \ourabstraction Abstraction}
\label{sec:recap_definition} 

We propose \ourabstraction to address the above challenges and
formalize early filtering opportunities for path queries with
list-based constraints through an automaton model that augments
an NFA with a dictionary. In contrast to previous work on similar automata
with memory, our goal is \textbf{not} to navigate the tradeoff between
expressiveness and complexity, but to \textbf{simplify specifying and
proving correctness of early filtering opportunities, as well as
enable syntactic conversion of the specification to an executable SQL query
that applies those early filters}. We will present a general construction
for any path query $Q = (S, R, \varphi)$ and illustrate it for $Q_B$.
For simplicity, we set $S = \{s\}$ (a single
start vertex $s$). It is straightforward to generalize our construction to
any set $S \subseteq V$.

\subsection{Regex-Only Case}\label{sec:nfa_in_sql}

\begin{figure}[tbp]
    \centering 
    \includegraphics[width=\linewidth]{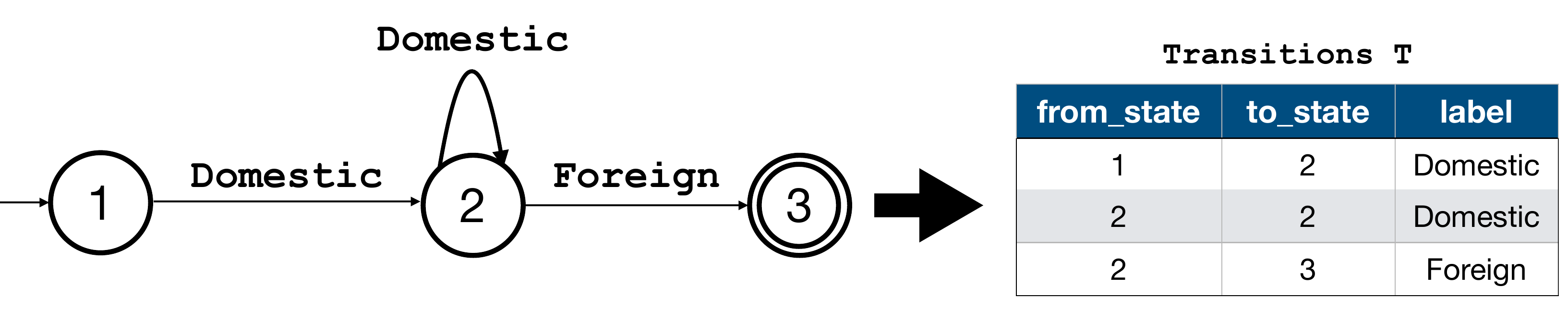}
    \caption{Left: NFA for regex $\texttt{Domestic}^+ \texttt{Foreign}$.
    Right: Tabular representation of the NFA.}
    \label{fig:NFA_as_table}
\end{figure}

Consider first a path query $Q = (S, R, \varphi = \true)$, i.e., one with
a label regex but without property constraints. $Q$'s \ourabstraction
is defined as the NFA for $R$. \Cref{fig:NFA_as_table} shows the NFA
for $Q_B$, together with its tabular representation $T$.
When $R$ is omitted, it is equivalent to \texttt{*}
and thus an NFA with a single state that is both start and accepting,
plus a self-loop.

\begin{figure}[tbp]
    \centering
    \begin{lstlisting}[language=SQL, escapechar=\%,
          basicstyle=\ttfamily\small,
          label={list:NFA_SQL}, caption={Semantics of a regex-only path query}]
    %{\textbf{WITH}}% RECURSIVE Paths AS (
        SELECT %{parameter:s}% as v,
               %{parameter:$q_0$}% as q
        UNION ALL
        SELECT E.dst as v, 
               T.to_state as q,
        FROM Paths P 
        JOIN Edges E ON P.v = E.src
        JOIN Transitions T ON T.from_state = P.q
        WHERE T.label = E.label
    )
    SELECT ... FROM Paths WHERE q IN %{parameter: $\mathfrak{Q}_F$}%
    \end{lstlisting}
\end{figure}

Of course, this NFA does not process strings, but paths
in input graph $G$. Hence, as in previous work on path queries
with regular expressions (see \Cref{sec:related}), the semantics
of \ourabstraction are defined by the product-graph
construction where the input graph is traversed jointly with the
NFA. (This also forms the foundation for the implementation
in SOTA systems that match paths based on label regex;
see \Cref{sec:related}).
The product-graph construction is equivalent to a join between
NFA and $G$, when representing the former as a transition table
(see \Cref{fig:NFA_as_table}) and the latter as a table of edge tuples.
We therefore present its formal semantics in SQL as a recursive join
using a Common Table Expression (CTE) as shown in \Cref{list:NFA_SQL}.
This has the advantage that we can directly execute this definition
in a relational DBMS.

The CTE defines a Paths table with columns $v$ (path vertex where the
next edge attaches) and $q$ (NFA state), initializing it with tuple
$(s, q_0)$, i.e., the start vertex of the path and the start state
of the NFA. These are passed as parameters to the SQL query.
(If there are multiple start states in $S$, then such a tuple is added
for each $s \in S$ by initializing Paths via a Cartesian product
of $S$ and $\{q_0\}$.)
The recursive member joins intermediate paths with the edge set,
thus expanding them by one more edge. Notice how the WHERE clause
enforces not only valid connections between graph edges, but also that a valid
NFA transition is taken. Non-determinism occurs when an intermediate
path can make multiple transitions with the same edge, i.e., multiple
transition tuples join with this path and edge.
Our approach supports \emph{any regex} over labels, generalizing those
supported by Cypher's \texttt{MATCH} clause.
The query that calls the CTE enforces that only paths whose label sequence
reaches an accepting state will be returned. The dots indicate that one
could return any information about the matched path, as long as that
information was collected by the recursive query.

\subsection{Property Constraints: Default Construction}
\label{sec:default_recap}

We now generalize the above construction to support any
property constraint $\varphi$, starting with a default construction.
The next section will extend it for early filtering on properties.

\begin{figure}[tbp]
\centering 
\includegraphics[width=\linewidth]{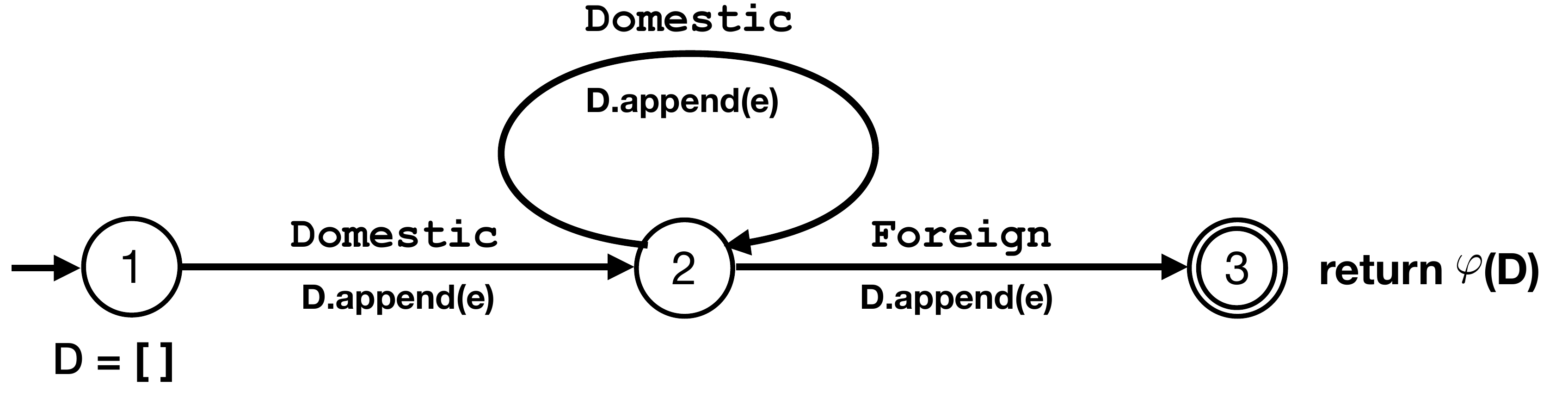}
\caption{Default construction: $\varphi$ is evaluated on the accumulated list $D$ only at an accepting state.}
\label{fig:naive_recap}
\end{figure}

The main idea is simple: Create the NFA for regex $R$ as before,
and maintain a dictionary $D$ for each path instance. It initially
holds the empty list. As edges are added to the path, they are
added to the list. When reaching an accepting
state, $D$ holds the data for all edges and hence we can evaluate
$\varphi$ on $D$ to determine if the property constraints are
satisfied by the path. \Cref{fig:naive_recap} illustrates this
construction for $Q_B$. Notice how edge transitions now also
append input edge $e$ that was used to make that transition in the
product-graph.
Unfortunately, this execution plan applies all list-based constraints
in the end, therefore it does not achieve early filtering on those constraints.
In short, it behaves like SOTA graph DBMS such as \NEO, \KUZU, and \MEMG,
which either resort to list processing or must pass intermediate paths to a different engine for post-processing
if they do not support list processing. This can waste huge amounts of resources:

\begin{example}[Wasted Computation]
\label{ex:wasted}
  Consider $Q_B$ on the graph in \Cref{fig:BitcoinGDB}, starting
  at Derrick's account.  Path
  $p = v_3 e_3 v_4 e_6 v_1 e_1 v_3 e_2 v_2 e_5 v_4$ (from
  Derrick to Hugo) matches the label regex, but fails property
  constraint $\varphi$.  Since its prefix
  $p' = v_3 e_3 v_4 e_6 v_1$ already fails the timestamp-range, meaning $p'$ is not viable (\Cref{def:doomed})
  and the resources spent on expanding it are wasted.
\end{example}

\subsection{Property Constraints: Early Filtering}
\label{sec:early_filtering}

The default construction, like the query plans used by SOTA graph DBMS,
applies list-based property constraints only after fully materializing
candidate paths. To eliminate non-viable paths early, we
must identify negatively stable constraints and maintain path state
that enables early detection when any of them becomes \false.
To this end, we extend the default construction with aggregate-style
functionality and predicates that verify viability early.

\begin{figure*}[tbp]
    \centering 
    \includegraphics[width=0.9\linewidth]{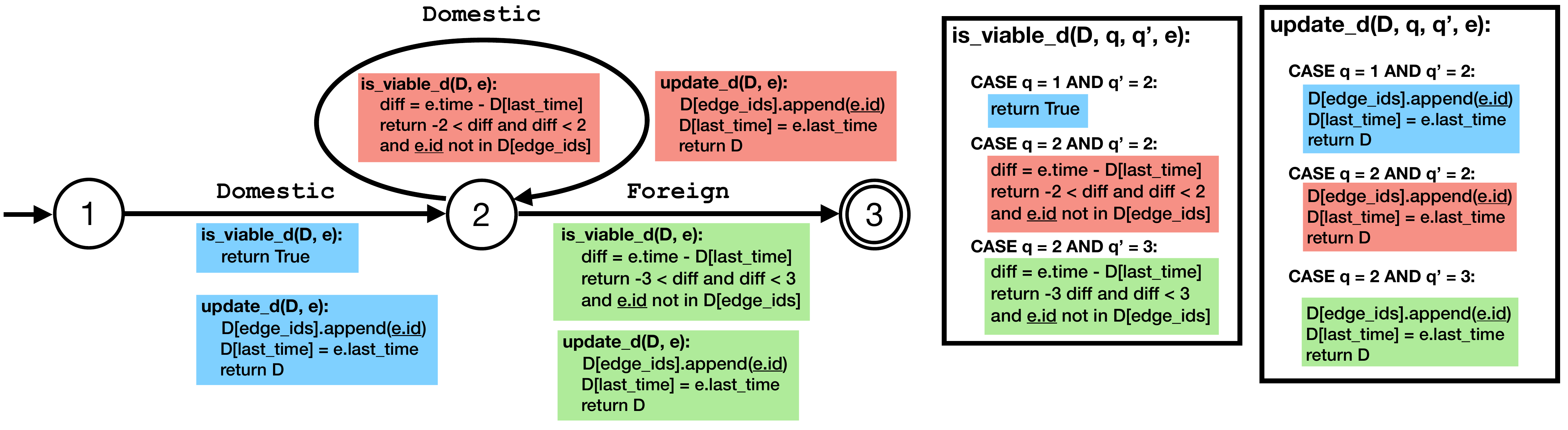}
    \caption{NFA with \isvalid and \update calls for $Q_B$.}
    \label{fig:full_recap}
\end{figure*}

Before formalizing this, consider the \ourabstraction for
$Q_B$ in \Cref{fig:full_recap}. As before, its structure is defined
by the NFA for $Q$'s regex $R$, augmented with a dictionary instance $D$
for each path candidate. But instead of a list of all edges,
$D$ now manages state specific to each negatively stable
constraint. For the trail semantics, this is the list of edge IDs;
for the time difference between consecutive transactions this is
$\mathtt{last\_time}$, the timestamp of the last transaction.
Note how the \update function associated with each state transition
performs the relevant updates on $D$: it appends the edge ID and
updates $\mathtt{last\_time}$. In addition to checking
the label, \isvalid checks for early filtering based on negative
stability. In the example, an intermediate path becomes non-viable
when the new edge $e$ causes a repeating edge ID or an out-of-range
timestamp.

While it is visually convenient to attach the relevant operations of
\update and \isvalid to each transition, each actually is a single function
with a CASE statement for the corresponding transition as
shown in \Cref{fig:full_recap}. We now formally define
\ourabstraction with
these and the other necessary functions to initialize and
finalize dictionary state and to enforce any leftover constraints
not enforced by \isvalid:

\begin{definition}[\ourabstraction]
\label{def:recap}
A \ourabstraction is a tuple
$(N,\allowbreak D,\allowbreak
\init, \update, \finalize, \isvalid,\allowbreak
\isvalidfinal,\allowbreak \merged, \ismergable)$ consisting of:
\begin{itemize}[noitemsep,nolistsep]
    \item An NFA $N = (\mathfrak{Q}, L, \delta, q_0, \mathfrak{Q}_F)$ (\Cref{def:NFA}).
    \item A dictionary $D$ of key-value pairs. A value is a scalar or a data structure, e.g., a list or a dictionary.
    \item Function $\init: \emptyset \rightarrow D$ to initialize the dictionary.
    \item Function $\update: D \times \mathfrak{Q} \times \mathfrak{Q} \times E \rightarrow D$ to update the dictionary.
    \item Function $\finalize: D \rightarrow D$ to convert the dictionary content to a new dictionary for returning it as the final result.
    \item Predicate $\isvalid: D \times \mathfrak{Q} \times \mathfrak{Q} \times E \rightarrow \{\true, \false\}$ to enforce dictionary-content-dependent constraints on state transitions.
    \item Predicate $\isvalidfinal: D \rightarrow \{\true, \false\}$ to enforce the final dictionary content satisfies the constraints.
    \item Function $\merged: D \times D \rightarrow D$ to merge two dictionaries.
    \item Predicate $\ismergable: D \times D \rightarrow \{\true, \false\}$ to enforce dictionary-content-dependent constraints when two dictionaries are merged.
\end{itemize}
\end{definition}

For example $Q_B$, \update and \isvalid are shown in \Cref{fig:full_recap}.
\init initializes $\mathtt{edge\_ids} = []$
and $\mathtt{last\_time}$ as NULL. The output of \finalize depends
on the desired query answer, e.g., it could return $\mathtt{edge\_ids}$
if one wants to know all edges belonging to a result path.
Here \isvalidfinal returns \true because all constraints are already
enforced by the NFA and \isvalid. However, if there was an additional
constraint that is not negatively stable, like ``the path must contain
two adjacent edges of the same color,'' then \isvalidfinal would
enforce that. The merge functions are only needed for advanced
query plans where paths are not assembled left-to-right, but
based on segments ``in the middle'' or bi-directional search.
To enable this, an additional key $\mathtt{first\_time}$ must
be added to $D$ to track the start time of a path segment.
Then, given dictionaries $D_1, D_2$, \merged returns a dictionary $D$
whose edge-ID list is the concatenation of $D_1.\mathtt{edge\_ids}$
and $D_2.\mathtt{edge\_ids}$. Similarly,
$D.\mathtt{first\_time} = D_1.\mathtt{first\_time}$ and
$D.\mathtt{last\_time} = D_2.\mathtt{last\_time}$.
\ismergable checks that
$D_1.\mathtt{edge\_ids} \cap D_2.\mathtt{edge\_ids} = \emptyset$ and
$D_1.\mathtt{last\_time} < D_2.\mathtt{first\_time}$.

The default construction (\Cref{sec:default_recap}) is a special case
of a selective aggregate where \init returns an empty list (implemented by
a dictionary with a single key whose value is the list), \update appends
the edge tuple to the list, \isvalid returns \true, and \isvalidfinal
evaluates $\varphi$ on the list.

\begin{figure}[tbp]
    \centering
    \begin{lstlisting}[language=SQL, escapechar=\%,
    basicstyle=\ttfamily\small,
    caption={SQL code defining any \ourabstraction as a simple recursive join query that calls the user-defined functions (underlined).},
    label={list:GeneralReCAP_SQL}]
    %{\textbf{WITH}}% RECURSIVE Paths AS (
        SELECT %{parameter: s}% AS v,
               %{parameter: $q_0$}% AS q,
               %{\underline{\init}}%() AS D
        UNION ALL 
        SELECT E.dst AS v,
               T.to_state AS q,
               %{\underline{\update}}%(P.D, P.q, T.to_state, E.*) AS D
        FROM Paths P
        JOIN Edges E ON P.v = E.src
        JOIN Transitions T ON P.q = T.from_state 
        WHERE T.label = E.label AND
            %{\underline{\isvalid}}%(P.D, P.q, T.to_state, E.*)
    )
    SELECT v, %{\underline{\finalize}}%(D)
    FROM Paths 
    WHERE q IN %{parameter: $\mathfrak{Q}_F$}% AND %{\underline{\isvalidfinal}}%(D)
    \end{lstlisting}
\end{figure}

The semantics of \ourabstraction are defined by the recursive SQL
query in \Cref{list:GeneralReCAP_SQL}. It is easy to show that
this produces exactly the paths that reach an accepting state in the NFA
and pass all \isvalid checks along the way. Hence the problem of declaring
opportunities for early filtering is reduced to defining the appropriate
dictionary state and \isvalid predicates.
In addition to providing a clean abstraction for specifying early filtering
opportunities via an NFA augmented with aggregate-style functionality,
a major benefit of \ourabstraction is that it does not require any
database engine modifications or SQL language extensions.
All customization for a given query
is encapsulated in the transitions table $T$ (for the regex $R$)
and a handful of functions (for property constraint $\varphi$).
Note how the WHERE clause of the recursive member enforces not only
the label pattern, but also enables early filtering via the \isvalid call.
The values of $s$, $q_0$, and $\mathfrak{Q}_F$ are passed as parameters.

To make this query executable in a relational DBMS, we encode
$D$ by default as a JSON document. When all its keys are associated with
types supported by the DBMS, then each dictionary key could be mapped
to a fresh column in Paths.

Notice that the query does not call \merged and \ismergable. That is
not surprising, because they are not needed to define the semantics.
These functions would be called ``behind the scenes,'' e.g., when
an advanced optimizer selects a plan that first finds
matches for pattern segments and then joins the corresponding path
segments together. While current relational optimizers do not support
this for recursive joins, \ourabstraction already provides an elegant
mechanism compatible with those strategies when they do. This is
analogous to parallel computation of distributive and algebraic
aggregates. Consider computing an average over a large set of numbers.
Different processors can each compute a (SUM, COUNT) pair for their
local data partition. When the aggregate comes with a merge
function (sum of individual sums, sum of individual counts),
then this function would not show up in the SQL query, but
a parallel DBMS optimizer can exploit its existence when
deciding how to best partition the problem and combine
the partial results.

\section{\ourabstraction Creation, Properties and Benefits}
\label{sec:recap_creation_benefits}

Having formally defined \ourabstraction, how does a user create the
specific instance for a given path query? And what are the tangible benefits
compared to not having this abstraction?

\subsection{Proving Correctness}
\label{subsec:LLMCorrectness}

To answer these questions, first consider the alternative.
The user may ask an advanced reasoning-capable LLM with a prompt like
``Rewrite the following recursive SQL query to make it run faster:
[text of query $Q$].'' Assume that the answer is indeed a rewritten
SQL query $Q'$ that ``looks right.'' When the result matters, 
the user needs a formal proof that $Q = Q'$.
This is generally undecidable, even for
linear recursion \cite{feder1992decidability}.
We tried the SOTA SQL equivalence provers
(SPES \cite{zhou2022spes}, WeTune \cite{wang2022wetune}, SQLSolver \cite{ding2023proving}), which either failed or did not support
recursive queries. When asking an advanced LLM to
propose a proof, the answer relied on arguments based on invariants
about maintained state of path prefixes. In short, the LLM
argued \emph{as if} the query can be abstracted as an
NFA-style structure with state managed along its transitions.
But it did not prove that the SQL text would indeed
yield such an abstraction.

\ourabstraction adds exactly this missing piece.
Instead of having to prove $Q = Q'$ for two long strings encoding
non-trivial SQL queries, \ourabstraction reduces the correctness
proof to 2 steps: First, prove that the general template in
\Cref{list:GeneralReCAP_SQL} produces exactly those paths
that reach an accepting state and pass all \isvalid and
\isvalidfinal checks. Second, prove that a path in $G$
reaches an accepting state and passes all \isvalid and
\isvalidfinal checks if and only if it is a query answer.
The first is relatively straightforward and needs to be done only
once, because the same template applies to all path queries.

The second must be done for each new query. But it is rather
mechanical as one proves invariants about the content of
dictionary $D$ and the corresponding validity checks
for each possible traversal sequence from start state $q_0$ to an
accepting state. This is trivial when there are no loops; loops
can be addressed via induction over path length.
Consider the \ourabstraction for $Q_B$
as defined in \Cref{sec:early_filtering}. There is only a single
traversal sequence from start state 1 to accepting state 3, and
it has a loop. Using induction over path length, it is easy to prove
that $D.\mathtt{edge\_ids}$ contains the IDs of those and only
those edges that were added to the path and that the value
in $D.\mathtt{last\_time}$ is the timestamp of the last edge.
Using this, it is easy to see that \isvalid will never eliminate
a query answer, while a non-answer will fail at least one of
these calls.

\subsection{Creating a \ourabstraction Instance}
\label{sec:creating_recap_instance}

How do we create a good \ourabstraction instance for a given path
query? The easiest approach is to use LLMs here as well, but
provide them with guardrails so that they do not produce a monolithic
rewritten query, but the functions and predicates of \ourabstraction.
(The NFA structure is easy to generate from the label regex using
standard constructions.) We had great success with
a prompt like ``For the query below, which finds paths in a graph that
satisfy certain conditions, create a representation as a finite state
machine with the following functions, all to be implemented as
DuckDB SQL functions: A function \init to initialize a dictionary that
holds relevant intermediate data, a function \update for each
state transition to update the dictionary, a function \isvalid
for each state transition to eliminate non-viable intermediate
results early based on dictionary content and a new edge processed,
and a function \isvalidfinal to check any remaining properties
of the path in the end. [query text]''

Receiving the code from the LLM response, the user can modify it
or start from scratch, relying on their own expertise and the examples
provided by our implementation (see \Cref{sec:recap_implementation}).
No matter how the \ourabstraction instance was created, in the
end its structure will always facilitate fairly mechanical
correctness proofs, either created manually or with the help of an LLM.
We next discuss a class of path queries for which both creation
and correctness proving are even simpler.

\subsection{Factorized \ourabstraction}
\label{sec:factorized}

Whenever the functions and predicates of a \ourabstraction instance
do not change their behavior depending on the NFA transition, this
instance is \emph{factorizable} into a standard NFA and what we refer
to as a \emph{selective aggregate}:

\begin{definition}[selective aggregate]
\label{def:selective_agg}
A \emph{selective aggregate} is a tuple
$(D, \init, \update, \finalize, \isvalid,\allowbreak
\isvalidfinal,\allowbreak \merged, \ismergable)$ consisting of:
\begin{itemize}[noitemsep,nolistsep]
    \item A dictionary $D$ of key-value pairs. A value is a scalar or a data structure, e.g., a list or a dictionary.
    \item Function $\init: \emptyset \rightarrow D$ to initialize the dictionary.
    \item Function $\update: D \times E \rightarrow D$ to update the dictionary.
    \item Function $\finalize: D \rightarrow D$ to convert the dictionary content to a new dictionary for returning it as the final result.
    \item Predicate $\isvalid: D \times E \rightarrow \{\true, \false\}$ to enforce dictionary-content-dependent constraints.
    \item Predicate $\isvalidfinal: D \rightarrow \{\true, \false\}$ to enforce the final dictionary content satisfies the constraints.
    \item Function $\merged: D \times D \rightarrow D$ to merge two dictionaries.
    \item Predicate $\ismergable: D \times D \rightarrow \{\true, \false\}$ to enforce dictionary-content-dependent constraints when two dictionaries are merged.
\end{itemize}
\end{definition}

When a \ourabstraction is factorizable, it eliminates the
``intertwining'' of the NFA state transitions with \update
and \isvalid in \Cref{def:recap}. Then the
\ourabstraction definition simplifies to a pair of an NFA
and a selective aggregate, where any NFA can be paired with
any (set of) selective aggregate(s). The correctness proof
simplifies to proving invariants for the selective aggregate
in isolation, without having to consider interactions
with any NFA structure. Hence we can maintain a library
of selective aggregates that are already proven correct for
certain constraints, allowing the user to pick and apply a suitable
one to the edge property column(s) of interest.
In that sense, a selective aggregate behaves like a standard UDA,
with the additional predicates to support early filtering.

Query $Q_A$ admits a factorizable \ourabstraction because it has no regex
and thus none of its property constraints depends on NFA transitions.
In $Q_B$, \isvalid changes behavior depending on the transition, which
prevents factorization. However, a careful analysis of $Q_B$
reveals that its \ourabstraction can be rewritten to
\emph{make it factorizable}.
Leaving all other functions and predicates as defined in
\Cref{sec:early_filtering}, we change \isvalid by replacing the
transition-dependent CASE conditions with equivalent conditions
that only use dictionary content and edge properties: Instead of
\texttt{CASE q=1 AND q'=2}, use
\texttt{CASE edge\_ids = []} because the edge-ID list is empty
exactly for the transition from state 1 to 2. Similarly, replace
\texttt{CASE q=2 AND q'=2} with 
\texttt{CASE edge\_ids $\neq$ [] AND e.label = Domestic} because
this property holds if and only if the automaton traverses the self-loop;
and similarly replace
\texttt{CASE q=2 AND q'=3} with 
\texttt{CASE edge\_ids $\neq$ [] AND e.label = Foreign}.
This construction works for this specific query, but other
\ourabstraction instances may not be factorizable.

\begin{figure}[tbp]
\centering
\includegraphics[width=\linewidth]{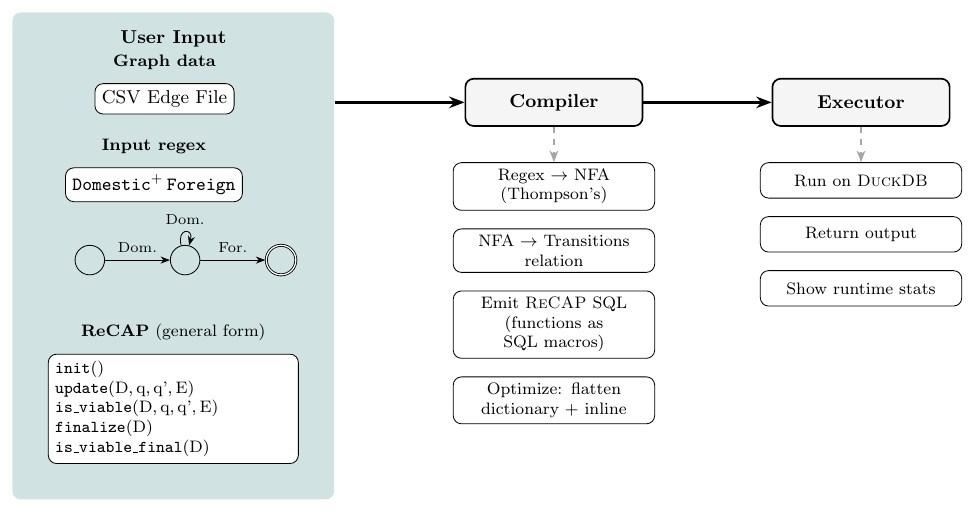}
\caption{\ourabstraction workbench pipeline. The author specifies a label
regex and the \ourabstraction functions and predicates. The Compiler
creates the recursive \ourabstraction SQL query
(\Cref{list:GeneralReCAP_SQL}) --- using \DUCKDB SQL macros and applies flattening.}
 \label{fig:recap_pipeline}
\end{figure}

\subsection{\ourabstraction Implementation}
\label{sec:recap_implementation}

We implemented a workbench tool (\Cref{fig:recap_pipeline}) that
aids the user in composing a \ourabstraction instance for their
path query (see \Cref{sec:exp_setup} for link to code repository). The steps are:
\begin{enumerate}
    \item Select a graph table in the database and declare start vertices,
    e.g., via ID list, a predicate over vertex properties, by out-degree
    quantile range, or every vertex. This will display sample rows and pre-fill later menus where columns can be selected for constraints. 
    \item Specify the (optional) label regex, which is automatically
    compiled into the Transitions relation using Thompson's construction (see \Cref{fig:NFA_as_table}).
    \item Additional form fields allow entering the 
    functions specifications of the \ourabstraction. For a full \ourabstraction, each row of the Transitions relation has 2 form fields to enter \update and \isvalid for each NFA transition. 
    \item  Upon clicking ``compile and run'' this specification is
    automatically compiled to an executable recursive SQL query
    following the template in \Cref{list:GeneralReCAP_SQL}.
    \item Finally, the output query is displayed alongside the runtime statistics and the output results of the Paths relation.
\end{enumerate}
For a factorized \ourabstraction, the workbench offers shortcuts
such as picking selective aggregates from a library.
In contrast to the general \ourabstraction, there is only
1 form field for \update and \isvalid, instead of one per transition.
Similarly, for users less familiar with SQL, we enable an optional text box
where they can explain the property constraint(s) in natural language.
An LLM (by default, Claude Sonnet 5 via the Anthropic API, or alternatively
a locally hosted Qwen2.5-7B-Instruct via Ollama) can then fill in the
\ourabstraction functions. The user is responsible for checking whether
these functions capture the semantics of their desired property constraint(s).
We plan to extend this interface to support different LLM API calls.
Each step comes with explanations and examples.

During the compilation step from user-provided \ourabstraction specification
to executable SQL, we also apply two low-level optimization steps.
\emph{Dictionary flattening} assigns each key of $D$ to its own
typed column of the \texttt{Paths} relation. For scalar types,
the corresponding SQL column type is used. When values are complex
objects, they are by default stored in the most appropriate
object type supported by the DBMS, e.g., a list, a struct, or a
JSON document. Similarly, \emph{function inlining}
inserts each function's body directly at its call site, using
column references to the flattened dictionary keys.
This completely eliminates costly packing and unpacking of data
into structs or JSON documents.
Without our optimizations, \DUCKDB can also expand macro calls
similar to function inlining, but it would still pack rows passed
to the macro into structs. Overall, our optimizations yield a consistent
1.1--2.1$\times$ runtime improvement and up to 2.5$\times$ memory
reduction across our benchmarks.

\subsection{Expressiveness}
\label{sec:expressiveness}

Previous work on generalizing path queries beyond regex on labels
(and their matching automaton models with memory) focused on
navigating the tradeoff between expressiveness and time complexity
(\Cref{sec:related}), showing that even small extensions can
quickly lead to NP hardness~\cite{gheerbrant2025dangers}.
\ourabstraction's goal is different: not to limit expressiveness,
but to \textbf{provide a general mechanism for exposing structure
that enables early filtering for all path queries, even those
with list-based property constraints.}
This mirrors SQL's evolution toward Turing-completeness:
rather than force the user to look elsewhere for a hard query, 
offer the most efficient implementation of it.

\begin{theorem}[Expressiveness]\label{thm:expressiveness}
\ourabstraction supports any constraint over both properties and
labels of edges.
\end{theorem}

It is trivial to add support for constraints involving
vertex labels and properties. To prove the theorem, encode the
entire constraint in $\varphi$ and apply the default construction (\Cref{sec:default_recap}) to an NFA with a single state (that is
start and accepting state) and a self-loop transition that collects
the list of edge data (label and properties). Clearly, this is not the
most efficient execution strategy. In particular, instead of including
a label regex in $\varphi$, moving it to the separate regex component $R$
of the path query will expose additional NFA states and transitions
where \isvalid can eliminate non-viable paths.
In general, exposing NFA structure and negatively stable constraints
as in \Cref{lem:conj_stable} will provide more opportunities for
early filtering. In the worst case, \ourabstraction behaves like SOTA systems:

\begin{observation}[Regret-Free Optimization]
\label{rem:regret-free}
When $\varphi$ is not negatively stable and cannot be
rewritten to apply \Cref{lem:conj_stable}, then the default
construction (\Cref{sec:default_recap}) produces a query plan
that reduces to the execution strategy of current SOTA systems:
apply early filtering using the label regex $R$, then check
property constraints on the complete path candidates.
\end{observation}

\subsection{Complexity Analysis}
\label{sec: complexity}

Since ``return all paths in $G$'' is a valid path query,
worst-case time complexity cannot be lower than the number of
possible paths $|\Pi(G)|$. This applies to queries that return actual paths, not just reachable vertices.
For more fine-grained results, we therefore
analyze complexity \emph{parameterized} on relevant factors: Let
$t_v$ denote the maximum time for a \isvalid or \isvalidfinal call,
$t_u$ the maximum time for a \update or \finalize call,
$\ell$ the maximum path length\footnote{It could be $\infty$
when edges and vertices are allowed to repeat and the graph has cycles.},
$P_\ell^{\mathrm{late}}$ the number of paths of length up to $\ell$
explored by the default construction, and
$P_\ell^{\mathrm{early}}$ the number explored under early filtering.
Since early filtering removes non-viable intermediate paths,
$P_\ell^{\mathrm{early}} \leq P_\ell^{\mathrm{late}}$, potentially
with a huge gap.

\begin{proposition}[Time Complexity]
\label{prop:runtime}
  The time complexity of \Cref{list:GeneralReCAP_SQL} is
  $O\left(P_\ell^{\mathrm{late}} \cdot (\ell + t_v + t_u)\right)$
  for the default construction and
  $O\left(P_\ell^{\mathrm{early}} \cdot \ell \cdot (t_v + t_u) \right)$
  for early filtering.
\end{proposition}

\begin{proof}
Consider a path $p$ of length up to $\ell$ explored by
\Cref{list:GeneralReCAP_SQL}. With the default construction,
each of $p$'s edges is appended to $p.D$, taking total time $\O(\ell)$.
In the end, \isvalidfinal is called and then \finalize to generate the
output tuple. With early filtering, each time an edge is added
to the path, \isvalid and \update are called. In the end,
\isvalidfinal and \finalize are called. We also need to account
for the join cost to assemble $p$. Follow standard practice, we
assume that with hash indices
on Edges (with \texttt{src} vertex as the key) and Transitions
(with combined key (\texttt{from\_state}, \texttt{label})),
the total join time is linear in the number of paths explored.
The corresponding query plan iterates through the Paths relation:
for each intermediate path, it looks up edges adjacent to
its end vertex in the Edges index and candidate transitions
in the Transitions index, producing the corresponding longer
intermediate paths.
\end{proof}

As illustrated by $Q_A$ and $Q_B$, the time complexity of
\ourabstraction functions and predicates is often linear
in dictionary size (or lower, e.g., constant time for appending
a new edge ID to \texttt{edge\_ids}):

\begin{corollary}
  If $t_v, t_u \in \O(\ell)$, then the time complexity of the
  default construction is $O\left(P_\ell^{\mathrm{late}} \cdot \ell\right)$.
  If $t_v, t_u \in \O(|D|)$ and dictionary size $|D|$ of the
  \ourabstraction is upper-bounded by a constant
  (i.e., independent of path length), then the time complexity for early
  filtering is $O\left(P_\ell^{\mathrm{early}} \cdot \ell \right)$.
\end{corollary}

\begin{proposition}[Space Complexity]
\label{prop:space}
  The space complexity of \Cref{list:GeneralReCAP_SQL} is
  $O(P_\ell^{\mathrm{late}} \cdot \ell)$ for the default construction
  and
  $O(P_\ell^{\mathrm{early}} \cdot \ell)$ for the early-filtering
  construction. When dictionary size $|D|$ of the selective aggregate
  is upper-bounded by a constant (i.e., independent of path length),
  then the latter bound tightens to
  $O(P_\ell^{\mathrm{early}})$.
\end{proposition}

\begin{proof}
  Each recursive tuple stores a prefix of length at most $\ell$,
  requiring $O(\ell)$ space.  The total is this per-tuple cost
  times the number of materialized prefixes.  Early filtering
  reduces the count from $P_\ell^{\mathrm{late}}$ to
  $P_\ell^{\mathrm{early}}$ but does not change the per-tuple
  cost.  In the early-filtering construction, if the query does not require the full path history
  (e.g., no trail semantics and no path output), and all
  constraints are maintainable with a bounded and constant summary (e.g., running sum, maximum and minimums, last timestamp seen), then
  per-tuple space can be reduced to $O(1)$, giving us a space complexity of 
  $O(P_\ell^{\mathrm{early}})$.
\end{proof}

\section{Experiments}
\label{sec:experiments}

We study \ourabstraction's effectiveness for early filtering on property
constraints, comparing it to SOTA graph and relational DBMS.

\subsection{Setup and Methodology}
\label{sec:exp_setup}

\begin{table}[tb]
\centering
\begin{tabular}{l l l}
\toprule
{Graph} &
{$|V|$}   &
{$|E|$}     \\
\midrule
\Bitcoin  &   5k & 35k \\
Metaverse &  1.3k & 78k \\
Reddit   & 35k  & 286k \\
LDBC100 (\texttt{P-K-P})  &   448k  & 19.9M \\
Datagen-7.7     &  13.1M  & 53.7M \\
\FinBench{} (SF10) & {1.0M} & {7.7M} \\
\bottomrule
\end{tabular}
\caption{Datasets used in the experiments.}
\label{tab:realdata}

\end{table}

\begin{table*}[tb]
\centering
\begin{minipage}[t]{\textwidth}
\centering
\begin{tabular}{l p{15.5cm}}
\toprule
\textbf{Query} & \textbf{Description} \\
\midrule
$Q_1$ & Label regex: $(\texttt{transfer} \mid \texttt{purchase} \mid \texttt{sale})^+ \cdot (\texttt{phishing} \mid \texttt{scam})^+$. Property constraints: chronological timestamps, same region, risk score range $\leq 20$ (in normal prefix), last normal edge risk $\geq 40$, total amount $\geq 1000$.\\
\midrule
$Q_2$ & Property constraint: there exist two adjacent edges that have the same color.\\
\midrule
$Q_3$ & Property constraint: strictly increasing numeric edge property (e.g., weight, timestamp). \\
\midrule
$Q_4$ & Property constraint: earliest and latest edge timestamp along the path does not exceed two weeks. \\
\bottomrule
\end{tabular}
\caption{Queries used in the experiments. All use trail semantics, i.e.,
no edge can be traversed more than once.}
\label{tab:queries}
\end{minipage}
\end{table*}

\textbf{Data and Queries.} The empirical evaluation relies on real
and synthetic data commonly used for path queries (\Cref{tab:realdata}),
obtained from
SNAP \cite{SPANData, BitcoinOTC, BitcoinOTC2, kumar2019predicting},
Kaggle \cite{kaggle_metaverse_transactions},
and the LDBC \FinBench generator \cite{qi2023ldbc}
with scale factor 10. For the latter, we only use 5 edge types:
\texttt{transfer}, \texttt{withdraw}, \texttt{deposit}, \texttt{signIn},
and \texttt{personOwnAccount}.
Due to a lack of established benchmarks for path queries with property constraints,
we create a mix of queries with varying opportunities for early filtering as summarized in \Cref{tab:queries}.

$Q_1$ analyzes financial data where directed edges represent transactions
with labels (\texttt{transfer}, \texttt{purchase}, \texttt{sale},
\texttt{phishing}, \texttt{scam}) and properties such as amount,
risk score, timestamp, and geographic region.
The query contains all types of constraints: label regex, negatively
stable constraints on individual edges and path aggregates, and
property constraints that are not negatively stable.
The regex specifies a transaction chain that looks normal at first,
then turns fraudulent. The property constraints ensure
(i) increasing timestamps for chronological event sequence \cite{yang2023anti},
(ii) a single jurisdiction (region),
(iii) a limited risk-score range for the ``normal'' initial transactions
to capture a consistently-profiled chain \cite{merklescience_risk_scores},
(iv) a high risk for the last ``normal'' transaction, reflecting its
role as the gateway to fraudulent activity 
\cite{chainalysis_japan_case_study_2023}, and
(v) a high total amount across the path to filter out financially
insignificant chains.

$Q_2$ has a property constraint that is not negatively stable:
the repeating color may appear on the very last edge of the path.
Since the given graphs do not have edge colors, we control
query selectivity through the number of distinct color values.
In the presented experiments, we add a color that is selected uniformly
at random from 5 possible values.
$Q_3$ is similar to $Q_1$, but only uses a monotonicity constraint
(which is negatively stable) along the path to enforce
strict temporal causality of events.
$Q_4$ enforces temporal cohesion (which is negatively stable)
across the path when connecting interactions in a social network.

For each query, we randomly selected start vertices from each
inter-quartile range of the out-degree distribution
(low: < 25\%, medium: 25--75\%, high: > 75\%) to include 
everything from ``poorly'' to ``well'' connected vertices.
No matter the start vertices, we observed the same trend.
Hence we show representative results with the following settings:
high-degree start vertices for $Q_1$,
medium-degree for $Q_2, Q_4$, and
low-degree for $Q_3$.

\textbf{Output and Success Measures.}
All queries collect the list of
edge IDs for each path and hence can return all result paths.
However, different graph and relational DBMS differ in how they
stage and return large outputs to the database client.
To eliminate this effect, we report the time until all paths have
been computed. This is achieved by having the DBMS return the path
count, which is a single number and thus eliminates the impact of
the transfer strategy from DBMS server to client.
In short, the reported times represent the time from query submission
until all paths have been fully materialized and counted.
In addition, we report the \emph{total} number of candidate paths
explored by each system. Without early filtering, this corresponds
to the number of paths matching regex $R$ with $\varphi = \true$.
For each query-graph pair, we search for paths of length up
to $\ell$ for $\ell \in \{2,\ldots,10\}$, running each configuration
5 times after one warm-up run. Results across runs were within 5\%
of each other and consistent for all systems, therefore we simply
report the median. A timeout of 2 hours was enforced per run.

\textbf{Systems and Environment.}
We compare \ourabstraction implemented via our workbench
(using \DUCKDB (v1.4.1)) against
\DUCKDB without \ourabstraction, and \SYSX, a SOTA commercial relational DBMS
known for its enterprise-level performance and reliability.
In addition, we compare against the leading graph DBMS
\NEO \cite{Neo4j} (v5.24.0, Community Edition),
\MEMG \cite{Memgraph} (v3.5.1), and
\KUZU \cite{feng2023kuzu} (v0.11.2).
For the relational DBMS, we create a standard adjacency index on the
Edges relation in advance, which is a hash index on the $\mathtt{src}$
column in \SYSX and an ART index in \DUCKDB. The graph DBMS maintain
an analogous index as well (a uniqueness constraint on
vertex ID for \NEO, a label-property index for \MEMG).
All experiments were executed on a server with a dual
Intel Xeon Gold 5218 CPU and 768 GB RAM. Our code and experimental setup is available as an open source repository\footnote{https://github.com/northeastern-datalab/ReCAP}.

\subsection{Running-Time Comparison}\label{sec:runtime_comparison}

\textbf{Summary.}
\ourabstraction outperforms \DUCKDB without \ourabstraction, \SYSX,
and the SOTA graph DBMS by orders of magnitude.
The gap widens as path length increases. As we show in 
\Cref{sec:total_int_results}, this advantage stems from early
filtering on list-based constraints, which reduces the number of
intermediate paths and enables \ourabstraction
to scale to paths roughly twice as long as the competition.

\begin{figure*}[tbp]
    \centering
 
    \begin{tikzpicture}
    \begin{axis}[
        name=plot1,
        xlabel={$\ell$},
        ylabel={Runtime (ms)},
        grid=both,
        width=0.28\textwidth,
        height=4.2cm,
        ymode=log,
        xtick={2,4,6,8, 10},
        title={$Q_1$ - Metaverse}
    ]
    \addplot[color=colReCAP, mark=o, thick] coordinates {
         (2, 23.7) (3, 34.5) (4, 58.0) (5, 76.7) (6, 108.6)
         (7, 124.3) (8, 146.7) (9, 158.5) (10, 164.9)
    };
    \addplot[color=colNeo4j, mark=diamond, thick] coordinates {
        (2, 387) (3, 17183) (4, 2405280)
    };
    \addplot[color=brown, mark=oplus, thick] coordinates {
        (2, 118) (3, 19311) (4, 2622685)
    };
    \addplot[color=colKuzu, mark=triangle, thick] coordinates {
        (2, 7786) (3, 992674)
    };
    \addplot[color=colDuckDB, mark=square, thick] coordinates {
        (2, 28) (3, 751) (4, 36698) 
    };
    \addplot[color=colSysX, mark=star, thick] coordinates {
        (2, 1631) (3, 115809) (4, 6066211) 
    };
    \draw[<->, thick, black] 
    (axis cs:4, 70) -- (axis cs:4, 22000)
    node[midway, right] {$\textbf{632}\times$};
 
    \end{axis}
    
    \begin{axis}[
        name=plot2,
        xlabel={$\ell$},
        grid=both,
        at={($(plot1.east)+(0.8cm,0)$)},
        anchor=west,
        width=0.28\textwidth,
        height=4.2cm,
        ymode=log,
        xtick={1,2,3,4,5},
        title={$Q_2$ - \Bitcoin}
    ]
 
    \addplot[color=colReCAP, mark=o, thick] coordinates {
         (2, 13.79) (3, 57.6) (4, 1780) (5, 88141)
    };
 
    \addplot[color=brown, mark=oplus, thick] coordinates {
        (2, 9) (3, 430) (4, 24572) (5, 1220810)
    };
    \addplot[color=colNeo4j, mark=diamond, thick] coordinates {
        (2, 297) (3, 668) (4, 22046) (5, 1323418) 
    };
    \addplot[color=colDuckDB, mark=square, thick] coordinates {
        (2, 51) (3, 272) (4, 12422) (5, 635708) 
    };
    \addplot[color=colSysX, mark=star, thick] coordinates {
        (2, 246) (3, 10435) (4, 608659)
    };
    
    \end{axis}
    
    \begin{axis}[
        name=plot3,
        at={($(plot2.east)+(0.8cm,0)$)},
        anchor=west,
        xlabel={$\ell$},
        grid=both,
        width=0.28\textwidth,
        height=4.2cm,
        ymode=log,
        xtick={2,4,6,8,10},
        title={$Q_3$ - Reddit}
    ]
    \addplot[color=colReCAP, mark=o, thick] coordinates {
        (2, 12.84) (3, 15.34) (4, 19.41) (5, 25.9) (6, 41.72) (7, 44.5) (8, 52.35) 
        (9, 53.19) (10, 64.05)
    };
    \addplot[color=colKuzu, mark=triangle, thick] coordinates {
        (2, 86) (3, 1054) (4, 194945)
    };
    \addplot[color=brown, mark=oplus, thick] coordinates {
        (2, 5) (3, 238) (4, 46769) 
    };
    \addplot[color=colNeo4j, mark=diamond, thick] coordinates {
        (2, 280) (3, 525) (4, 50168)
    };
    \addplot[color=colDuckDB, mark=square, thick] coordinates {
        (2, 10) (3, 127) (4, 21454) (5, 6018242) 
    };
    \addplot[color=colSysX, mark=star, thick] coordinates {
        (2, 22) (3, 9644) (4, 1697847) 
    };
 
    \draw[<->, thick, black] 
    (axis cs:5,20) -- (axis cs:5,4018242)
    node[midway, right] {$\textbf{232,364}\times$};
    \end{axis}
    
    \begin{axis}[
        name=plot4,
        at={($(plot3.east)+(0.8cm,0)$)},
        anchor=west,
        xlabel={$\ell$},
        grid=both,
        width=0.28\textwidth,
        height=4.2cm,
        ymode=log,
        xtick={2,4,6,8,10},
        title={$Q_4$ - LDBC100}
    ]
 
    \addplot[color=colReCAP, mark=o, thick] coordinates {
        (2, 37.76) (3, 48.07) (4, 90.62) (5, 134.70)
        (6, 433.87) (7, 2146.80) (8, 12408.70)
    };
    \addplot[color=colKuzu, mark=triangle, thick] coordinates {
        (2, 1736.87) (3, 2355.70) (4, 30136.91) (5, 1636342.13)
    };
    \addplot[color=colNeo4j, mark=diamond, thick] coordinates {
        (2, 445.83) (3, 676.03) (4, 13916.86) (5, 928776.36)
    };
    \addplot[color=colDuckDB, mark=square, thick] coordinates {
        (2, 68) (3, 148) (4, 4525) (5, 223916) 
    };
    \addplot[color=colMemgraph, mark=oplus, thick] coordinates {
        (2, 4.07) (3, 252.57) (4, 15488.84) (5, 881075.28)
    };
    \addplot[color=colSysX, mark=star, thick] coordinates {
        (2, 426) (3, 9899) (4, 580378) 
    };
    
    \draw[<->, thick, black] 
    (axis cs:5,200) -- (axis cs:5,92000)
    node[midway, right, yshift=4mm] {$\textbf{1,671}\times$};
    \end{axis}
    \end{tikzpicture}
    
    \caption{Running-time comparison across queries and graphs.
    \textcolor{colReCAP}{$\circ$}~\ourabstraction,
    \textcolor{colKuzu}{$\triangle$}~\KUZU,
    \textcolor{colNeo4j}{$\diamond$}~\NEO,
    \textcolor{brown}{$\oplus$}~\MEMG,
    \textcolor{colDuckDB}{$\square$}~\DUCKDB,
    \textcolor{colSysX}{$\star$}~\SYSX.
    Competitors scale poorly to larger $\ell$ due to huge intermediate results;
    \ourabstraction pushes property constraints deep into path exploration and
    scales to more than double the path length. The exception is $Q_2$, which does
    not admit early filtering.}
    \label{fig:performance_grid}
\end{figure*}

Running $Q_1$ in \NEO, \MEMG, and \KUZU exposes two inefficiencies.
First, limited regex support forces decomposition into two segments
joined at an intermediate vertex, inducing a plan where matches
for each segment are found separately and then joined.
Second, all property constraints are applied at the end, yielding huge
intermediate results: for $\ell=4$, the normal prefix alone produces
over $25M$ candidate paths, of which only $5k$ are answers.
In contrast, \ourabstraction maintains path state
(timestamp, region, risk range, last risk, cumulative amount),
filtering non-viable paths early. As \Cref{fig:performance_grid} shows,
this yields for $\ell=4$ a 41,470$\times$ speedup over the graph DBMS
and 632$\times$ over \DUCKDB without \ourabstraction.
The advantage of \DUCKDB without \ourabstraction over the graph DBMS
comes from \emph{our implementation of the regex} as a join with the
transitions table. Nonetheless, all competitors share the same
late-filtering approach for property constraints. \SYSX, despite
also using a join for the regex, is further penalized by its lack
of native array types, requiring management of JSON-encoded path state.
 
Since $Q_2$ does not admit early filtering, all systems share the same
query plan. Yet \ourabstraction still achieves a speedup for larger outputs
by piggybacking color checks on path exploration rather than doing them
in a separate post-processing pass. Specifically, \ourabstraction tracks
\texttt{last\_color} and a Boolean \texttt{completed}, checked in
\isvalidfinal. For $\ell=5$, this yields a 12$\times$ speedup over the
graph DBMS and 6$\times$ over \DUCKDB without \ourabstraction,
illustrating that incremental constraint evaluation reduces overhead
even when early filtering is not possible. $Q_3$ and $Q_4$, like $Q_1$, admit early filtering. \Cref{fig:performance_grid} shows competitors timing out at $\ell \le 6$; at $\ell=5$, \ourabstraction is 232,364$\times$ and 1,671$\times$ faster than the best competitor \DUCKDB.

\subsection{Automata vs. Property Filtering}
\label{subsec:e4_isolation}
 
Compiling a label regex into an automaton is already known to improve
running time over checking labels only at the end. To confirm \ourabstraction's
gains go beyond this, we hold regex processing fixed and vary only how the
list-based property constraint is checked, on $Q_1$-Metaverse: (1)~\textbf{Regex-only}
(constraint removed); (2)~\textbf{Regex + late} (constraint checked at the end on
complete paths, as in SOTA systems); (3)~\textbf{Regex + early} (full
\ourabstraction).

\begin{table}[tb]
\centering
\small
\begin{tabular}{l l r r r}
\toprule
$\ell$ & Configuration & \# Results & \# Interm.\ rows & Runtime (ms) \\
\midrule
2 & 1. Regex-only            & 1{,}408   & 15{,}934     & 16.0 \\
2 & 2. Regex + late          & 23        & 15{,}934     & 27.5 \\
2 & 3. Regex + early   & 23        & 65     & 23.2 \\
\midrule
3 & 1. Regex-only            & 69{,}500  & 790{,}814    & 164.8 \\
3 & 2. Regex + late          & 95        & 790{,}814    & 273.9 \\
3 & 3. Regex + early   & 95        & 149      & 41.5 \\
\midrule
4 & 1. Regex-only            & 3{,}530{,}274 & 35{,}384{,}382 & 6{,}466.7 \\
4 & 2. Regex + late          & 264       & 35{,}384{,}382 & 11{,}798.2 \\
4 & 3. Regex + early   & 264       & 328      & 67.6 \\
\bottomrule
\end{tabular}
\caption{Isolating automata exploration from property early-filtering on
$Q_1$-Metaverse. Configs 1--2 share intermediate paths (differ in whether the
constraint is enforced); configs 2--3 share output (differ in when it is checked).}
\label{tab:e4_isolation}
\end{table}

As \Cref{tab:e4_isolation} demonstrates, regex-only wins by about a factor
of 2 compared to regex + late. This is expected, as essentially all
intermediate paths are scanned to check the property constraints.
Regex + early always wins over regex + late, but it also dramatically
improves over regex-only for selective property constraints that are
negatively stable. Here, the additional cost for calling
\update and \isvalid during NFA transitions is more than compensated
by the hugely reduced number of intermediate results.

\begin{figure*}[tbp]
    \centering
    
    \begin{tikzpicture}
 
    \begin{axis}[
        name=plot1,
        anchor=west,
        xlabel={$\ell$},
        ylabel={Total Intermediate Results},
        grid=both,
        width=0.28\textwidth,
        height=4.2cm,
        ymode=log,
        xtick={2,4,6,8, 10},
        title={$Q_1$ - Metaverse}
    ]
    \addplot[color=colReCAP, mark=o, thick] coordinates {
        (2, 65) (3, 149) (4, 328) (5, 532) (6, 777) (7, 811)
        (8, 906) (9, 918) (10, 944) 
    };
    \addplot[color=orange, mark=square, thick] coordinates {
        (2, 1408) (3, 69500) (4, 3530598)
    };
    \addplot[color=colNeo4j, mark=triangle, thick] coordinates {
        (2, 1408) (3, 69500) (4, 3530274)
    };
 
    \draw[<->, thick, black] 
    (axis cs:4,500) -- (axis cs:4,2630274)
    node[midway, right, yshift=4mm] {$\textbf{10,763}\times$};
 
    \end{axis}
 
    \begin{axis}[
        name=plot2,
        at={($(plot1.east)+(0.8cm,0)$)},
        anchor=west,
        xlabel={$\ell$},
        grid=both,
        width=0.28\textwidth,
        height=4.2cm,
        ymode=log,
        xtick={2,3,4,5},
        title={$Q_2$ - Bitcoin}
    ]
    \addplot[color=colReCAP, mark=o, thick] coordinates {
        (2, 3862) (3, 163710) (4, 8263449) (5, 372223533)
    };
    \addplot[color=orange, mark=square, thick] coordinates {
        (2, 3862) (3, 163756) (4, 8270764) (5, 372750847)
    };
    \addplot[color=colNeo4j, mark=triangle, thick] coordinates {
        (2, 3862) (3, 163710) (4, 8263449) (5, 372223533)
    };
    \end{axis}

    \begin{axis}[
        name=plot3,
        at={($(plot2.east)+(0.8cm,0)$)},
        anchor=west,
        xlabel={$\ell$},
        grid=both,
        width=0.28\textwidth,
        height=4.2cm,
        ymode=log,
        xtick={2,4,6,8, 10},
        title={$Q_3$ - Reddit}
    ]
    \addplot[color=colReCAP, mark=o, thick] coordinates {
        (2, 68) (3, 254) (4, 1160) (5, 4792) (6, 16674) (7, 49384)
        (8, 128854) (9, 299428) (10, 627472) 
    };
    \addplot[color=orange, mark=square, thick] coordinates {
        (2, 287) (3, 125772) (4, 20876598) (5, 4792152245)
    };
    \addplot[color=colNeo4j, mark=triangle, thick] coordinates {
        (2, 287) (3, 125772) (4, 20873911)
    };
        \draw[<->, thick, black] 
    (axis cs:5,7000) -- (axis cs:5,2900000000)
    node[midway, right, yshift=4mm] {$\textbf{1,000,031}\times$};
    \end{axis}
    
    \begin{axis}[
        name=plot4,
        at={($(plot3.east)+(0.8cm,0)$)},
        anchor=west,
        xlabel={$\ell$},
        grid=both,
        width=0.28\textwidth,
        height=4.2cm,
        ymode=log,
        xtick={2,4,6,8,10},
        title={$Q_4$ - LDBC100}
    ]
    \addplot[color=colReCAP, mark=o, thick] coordinates {
        (2, 422) (3, 4621) (4, 38444) (5, 279035)
        (6, 1818120) (7, 10704627) (8, 57615956)
    };
    \addplot[color=colKuzu, mark=triangle, thick] coordinates {
        (2, 1783) (3, 107611) (4, 5637415) (5, 277974425)
    };
    \addplot[color=colNeo4j, mark=diamond, thick] coordinates {
        (2, 1783) (3, 107611) (4, 5637415) (5, 277974425)
    };
    \addplot[color=colMemgraph, mark=oplus, thick] coordinates {
        (2, 1783) (3, 107611) (4, 5637415) (5, 277974425)
    };
 
    \draw[<->, thick, black] 
    (axis cs:5,350000) -- (axis cs:5,220974425)
    node[midway, right, yshift=3mm] {$\textbf{996}\times$};
    \end{axis}
    \end{tikzpicture}
    
    \caption{Total intermediate paths generated across queries and datasets.
    \textcolor{colReCAP}{$\circ$}~\ourabstraction,
    \textcolor{orange}{$\square$}~\KUZU/\SYSX/\DUCKDB,
    \textcolor{colNeo4j}{$\triangle$}~\NEO/\MEMG.
    \NEO/\MEMG\ enforce trail semantics during matching; \KUZU/\SYSX/\DUCKDB\ use
    walk semantics and enforce trail afterward, so each group produces the same count.}
    \label{fig:intermediate_total_grid}
\end{figure*}

\subsection{Intermediate Result Size Comparison}
\label{sec:total_int_results}
 
To understand the root cause of \ourabstraction's running-time improvement,
we report intermediate path counts in \Cref{fig:intermediate_total_grid}.
It is clear that intermediate
result cardinality closely mirrors the running-time trend.
Since the difference between \ourabstraction and plain \DUCKDB is early
filtering, intermediate cardinality directly explains the running-time gap
between them. By inspecting query plans, we confirmed that all competitors
first generate candidate paths based solely on the label regex (if specified)
and then post-process them with the property constraints. \ourabstraction
changes this behavior for \DUCKDB by pushing property constraints deep into
path exploration. The graph DBMS currently offer no comparable mechanism:
In their \texttt{MATCH} clause, Cypher, GQL, and SQL/PGQ cannot express
list-based property constraints like those in our queries
\cite{gheerbrant2024gql, gheerbrant2025dangers}, forcing evaluation
over complete paths via ``list processing.''
While \NEO and \MEMG enforce trail semantics (no repeated edge) during the
regex matching phase, \KUZU, \SYSX and \DUCKDB enforce it afterward, applying
walk semantics initially (which allows repeat vertices and edges).
As \Cref{fig:intermediate_total_grid}
shows, the cardinality difference between early and delayed enforcement
of trail semantics is negligible here. The reasons are (1) sparseness of the graph,
(2) small path length, and, for $Q_1$,
(3) the selectivity of the regex that makes loops almost impossible
for short paths.

For $Q_1$, at $\ell=4$, the competitors generate over \emph{3.5 million} intermediate
paths, whereas \ourabstraction produces only $324$, which
is remarkably close to the final output of $264$.
Since $Q_2$ admits no early pruning, all systems produce the same intermediate paths. The trail vs.\ walk distinction is again negligible: at $\ell=3$, trail-enforcing systems produce 163,710 paths vs. 163,756 for the others; at $\ell=5$, it is 372,223,533 vs. 372,750,847.
For $Q_3$ and $Q_4$, the pattern mirrors $Q_1$. At $\ell=4$, \ourabstraction generates $307$ intermediate paths for $Q_3$, versus 20,873,911 for \NEO and \MEMG and 20,876,598 for \KUZU, \SYSX and \DUCKDB; at $\ell=5$, the competitors produce over a million times as many! For $Q_4$, the competitors generate 5,637,415 and 277,974,425 intermediate paths at $\ell=4$ and $\ell=5$, respectively, compared to 38,444 and 279,035 for \ourabstraction.

\subsection{Peak Memory Comparison}
\label{subsec:memory_comparison}

\begin{table}[tb]
\centering
\small
\begin{tabular}{r rrrr}
\toprule
$\ell$ & \KUZU & \NEO & \MEMG & \ourabstraction \\
\midrule
2 & 6{,}255   & 16{,}044 & 5{,}979 & 471 \\
3 & 6{,}403   & 16{,}049 & 5{,}981 & 464 \\
4 & 11{,}575  & 16{,}060 & 5{,}983 & 508 \\
5 & 322{,}385 & 16{,}083 & 5{,}982 & 4{,}073 \\
\bottomrule
\end{tabular}
\caption{Peak memory (MB) on $Q_4$ (LDBC100). \NEO/\MEMG\ stay flat (JVM heap reservation); \KUZU's
memory grows $28\times$ from $\ell=4$ to $\ell=5$ versus \ourabstraction's
$8\times$.}
\label{tab:memory_q4}
\end{table}

Peak memory complements the running-time story: \ourabstraction
uses the least memory (hundreds of MB to a few GB) across every query
and $\ell$ tested, even where intermediate cardinality reaches tens
of millions of rows. The numbers for \NEO and \MEMG reflect JVM
heap management and resident graph storage, not query-specific state.
\KUZU's memory scales with intermediate cardinality the same way its
running time does, from a few GB up to \textbf{322\,GB} on $Q_4$ at
$\ell=5$ --- roughly $80\times$ \ourabstraction's footprint.

\subsection{Evaluation on a Standardized Benchmark}
\label{subsec:finbench}

We evaluate \ourabstraction on 3 transaction-chain
queries (\texttt{TCR1}, \texttt{TCR5}, \texttt{TCR8}) from the
\FinBench benchmark's reference implementation.
All three share the pattern that motivated this paper: a bounded-hop
chain with a per-hop property constraint. We refer the reader to \cite{qi2023ldbc} for query specifics.
We evaluate at SF10, produced by the LDBC \FinBench
generator 
and loaded directly from
the raw CSV output. \ourabstraction is created via our workbench,
while \NEO, \MEMG, and \KUZU each run their Cypher translation.

\Cref{fig:finbench_competitors} reports \ourabstraction vs.\
\NEO, \MEMG, and \KUZU. \ourabstraction completes
the full range on all three queries. Each
competitor is plotted as far as it actually completes: \KUZU\ exhausts its
buffer pool by $\ell=7$ on \texttt{TCR1} and \texttt{TCR5}; on
\texttt{TCR8}, all three competitors fail by $\ell=8$ (\NEO, \MEMG time
out, \KUZU runs out of memory) while \ourabstraction alone
completes it.
On \texttt{TCR1}, running times are similar until $\ell=4$, where few
intermediate paths are eliminated beyond the regex; the gap widens once
early filtering can make a difference (by $\ell=6$, 18.0M intermediate paths vs.\
1.2M final results).

At the farthest point every remaining competitor reaches, the gap is
already large and still widening: $490$--$783\times$ on \texttt{TCR1}
($\ell=8$, \NEO/\MEMG), $286$--$421\times$ on \texttt{TCR5} ($\ell=8$, \NEO/\MEMG). \KUZU's behavior---slower than \NEO/\MEMG\ at small $\ell$, with the gap narrowing as $\ell$ grows---stems from a fixed cost that scales with graph size rather than with $\ell$ or selectivity, which \NEO/\MEMG's index-free adjacency avoids. The effect is not a measurement artifact: a discarded warmup run leaves the $\ell=2$ time essentially unchanged, so the cost is executed on every run, not incurred at compilation.
\begin{figure*}[tbp]
    \centering
    \begin{tikzpicture}
    \begin{axis}[
        name=fbplot1,
        xlabel={$\ell$},
        ylabel={Runtime (ms)},
        grid=both,
        width=0.31\textwidth,
        height=4.2cm,
        ymode=log,
        xtick={2,3,4,5,6,7,8},
        title={\texttt{TCR1}}
    ]
    \addplot[color=colReCAP, mark=o, thick] coordinates {
        (2,21.9) (3,207.9) (4,1019.7) (5,3282.2) (6,5533.1) (7,6325.9) (8,6350.0)
    };
    \addplot[color=colNeo4j, mark=diamond, thick] coordinates {
        (2,25.8) (3,98.1) (4,4636.9) (5,49827.9) (6,384545.7) (7,1870044.7) (8,4973949.3)
    };
    \addplot[color=colMemgraph, mark=oplus, thick] coordinates {
        (2,15.3) (3,62.4) (4,2942.0) (5,36224.5) (6,246415.5) (7,1173741.3) (8,3109580.9)
    };
    \addplot[color=colKuzu, mark=triangle, thick] coordinates {
        (2,54.0) (3,68.1) (4,608.0) (5,9449.5) (6,72580.4)
    };
    \end{axis}
 
    \begin{axis}[
        name=fbplot2,
        at={($(fbplot1.east)+(0.8cm,0)$)},
        anchor=west,
        xlabel={$\ell$},
        grid=both,
        width=0.31\textwidth,
        height=4.2cm,
        ymode=log,
        xtick={2,3,4,5,6,7,8},
        title={\texttt{TCR5}}
    ]
    \addplot[color=colReCAP, mark=o, thick] coordinates {
        (2,13.7) (3,29.3) (4,55.4) (5,81.6) (6,131.2) (7,177.6) (8,188.1)
    };
    \addplot[color=colNeo4j, mark=diamond, thick] coordinates {
        (2,18.2) (3,86.4) (4,452.0) (5,1994.1) (6,7724.4) (7,22761.4) (8,53733.9)
    };
    \addplot[color=colMemgraph, mark=oplus, thick] coordinates {
        (2,15.3) (3,88.9) (4,523.1) (5,2476.1) (6,10007.0) (7,31562.7) (8,79137.6)
    };
    \addplot[color=colKuzu, mark=triangle, thick] coordinates {
        (2,38279.6) (3, 54578.2) (4,85772.8) (5,188131.8) (6,686617.8)
    };
    \end{axis}
 
    \begin{axis}[
        name=fbplot3,
        at={($(fbplot2.east)+(0.8cm,0)$)},
        anchor=west,
        xlabel={$\ell$},
        grid=both,
        width=0.31\textwidth,
        height=4.2cm,
        ymode=log,
        xtick={2,3,4,5,6,7,8},
        title={\texttt{TCR8}}
    ]
    \addplot[color=colReCAP, mark=o, thick] coordinates {
        (2,19.2) (3,62.8) (4,151.9) (5,290.7) (6,448.2) (7,685.3) (8,928.3)
    };
    \addplot[color=colNeo4j, mark=diamond, thick] coordinates {
        (2,64.2) (3,835.0) (4,5704.3) (5,43939.7) (6,366497.9) (7,3093516.2)
    };
    \addplot[color=colMemgraph, mark=oplus, thick] coordinates {
        (2,54.0) (3,913.4) (4,8090.0) (5,78488.7) (6,669250.7) (7,6077497.6)
    };
    \addplot[color=colKuzu, mark=triangle, thick] coordinates {
        (2, 55308.7) (3, 90708.0) (4,261065.5) (5,552027.6) (6,1137635.8) (7,2734055.5)
    };
    \end{axis}
    \end{tikzpicture}
 
    \caption{Running time on \FinBench{} SF10,
    \texttt{TCR1}/\texttt{TCR5}/\texttt{TCR8}, $\ell=2..8$:
    \textcolor{colReCAP}{$\circ$}~\ourabstraction,
    \textcolor{colNeo4j}{$\diamond$}~\NEO,
    \textcolor{colMemgraph}{$\oplus$}~\MEMG,
    \textcolor{colKuzu}{$\triangle$}~\KUZU.
    \KUZU\ exhausts memory on \texttt{TCR1}/\texttt{TCR5}; on \texttt{TCR8} every
    competitor fails by $\ell=8$ while \ourabstraction alone completes it.}
    \label{fig:finbench_competitors}
\end{figure*}

\subsection{Segmented Path Exploration}
\label{subsec:r4_o2_compatibility}

While current relational DBMS implement recursive joins using the
default left-deep plan, future systems could consider more advanced plans
that split the path pattern into segments and apply bi-directional
search \cite{nguyen2021efficiently,KoschRPQsBG,TRE_RPQ,yakovets2015waveguide}.
An abstraction like \ourabstraction plays an essential role for enabling
users to specify how state for separately matched path segments
can be maintained independently and then merged as needed.
\ourabstraction achieves this through the \merged and \ismergable functionality
(\Cref{def:recap}).

To demonstrate the potential for performance improvements, we compare
running time of a variant $Q'_1$ of $Q_1$ (whose only property constraint
is timestamp monotonicity) when using the standard left-deep plan vs. a
more advanced plan that splits the pattern in half. Concretely,
$Q_1$'s regex is split into $\mathtt{(transfer|purchase|sale)+}$ and
$\mathtt{(phishing|scam)+}$. Since \DUCKDB only supports left-deep
evaluation of recursive joins, we force the advanced plan as follows:
Create separate \ourabstraction instances for each half of the
pattern (which consist of the corresponding regex part and
the monotonicity constraint). After computing the path segments
for each half separately, we join them on their respective
end vertices while also applying \merged and \ismergable.
(Each segment keeps track of both the first and the last
vertex, as well as the first and the last timestamp.)

\begin{table}[tb]
\centering
\small
\sisetup{table-format=1.2e1, exponent-product=\times}
\begin{tabular}{r r S[table-format=6.0] S[table-format=6.0] r}
\toprule
$\ell$ & \#Paths & {Left-deep (ms)} & {Split (ms)} & Speedup \\
\midrule
2 & 719             & \num{1.06e1} & \num{1.89e1} & 0.6$\times$ \\
3 & 8{,}440         & \num{2.57e1} & \num{3.97e1} & 0.6$\times$ \\
4 & 125{,}267       & \num{1.82e2} & \num{1.23e2} & 1.5$\times$ \\
5 & 1{,}201{,}453   & \num{2.42e3} & \num{4.49e2} & 5.4$\times$ \\
6 & 9{,}241{,}189   & \num{2.18e4} & \num{3.49e3} & 6.2$\times$ \\
7 & 61{,}430{,}146  & \num{1.89e5} & \num{2.88e4} & 6.6$\times$ \\
8 & 394{,}855{,}868 & \num{2.24e6} & \num{2.95e5} & 7.6$\times$ \\
\bottomrule
\end{tabular}
\caption{Left-deep vs.\ advanced split query plan for $Q'_1$
(same fixed start vertex, running time in ms).}
\label{tab:segmented_split}
\end{table}

As \Cref{tab:segmented_split} shows, for small $\ell$ the advanced plan
is slightly slower, but it starts winning at $\ell \ge 4$,
delivering almost an order of magnitude improvement at $\ell=8$.
Intuitively, here the split plan wins because each half explores
shorter path fragments. (Since each half consists of
at least 1 edge, the split plan explores segments of length up to
$\ell-1$ per half, while the left-deep plan explores up to length
$\ell$.) Note that we do not claim general superiority of split plans,
but only that they can offer significant benefits and that they are easy
to enable with an abstraction like \ourabstraction. Without \ourabstraction,
it is not clear where one would declare \merged and \ismergable,
because they are not part of the query logic itself.

\subsection{Scaling to Larger Graphs}\label{subsec:scalability_challenge}

The above results show how the combinatorial nature of path queries
that generate all matching paths---instead of only reporting
the set of reachable end vertices---can cause huge intermediate and final
results even on small graphs. Of course, for sufficiently selective
constraints, efficient approaches like \ourabstraction can scale
to larger graphs as the following experiment demonstrates.
Here we evaluate on Datagen-7.7 (53.7M edges) an aggregate constraint
where the product of individual edge weights along a path is
upper bounded by parameter $U$.
It is negatively stable since all weights fall in range $[1.0,1.75]$:
the product never decreases and hence cannot recover once it exceeds $U$.
Since the product's floor ($1.0^\ell$) does not grow with $\ell$, the
constraint does not become trivially unsatisfiable at larger $\ell$.
We use the dataset's maximum-out-degree vertex as the start vertex and
set $U=1.05$. The comparison only includes \MEMG, the fastest SOTA
graph DBMS on this problem.
\Cref{tab:scalability_challenge} summarizes the results.

\begin{table}[tb]
\centering
\small
\begin{tabular}{r r rr r}
\toprule
$\ell$ & \#Results & \ourabstraction & \MEMG & Speedup of \ourabstraction\\
\midrule
2 & $15{,}834$        & $0.05$\,s   & $0.06$\,s    & $1.3\times$ \\
3 & $2{,}946{,}070$    & $0.62$\,s   & $8.05$\,s    & $13.0\times$ \\
4 & $57{,}261{,}675$   & $5.39$\,s   & $322.7$\,s   & $59.9\times$ \\
5 & $5{,}723{,}628{,}083$ & $869.2$\,s  & TO & --- \\
\bottomrule
\end{tabular}
\caption{$\prod\text{weight}\le 1.05$ on Datagen-7.7, start vertex
$6597072984304$. \#Results is the cumulative number of result paths of
length $2$ to $\ell$; \ourabstraction\ and \MEMG\ return identical counts at
every $\ell$ both complete. TO: timeout after 30\,min.}
\label{tab:scalability_challenge}
\end{table}

Due to the combinatorial nature of the query, the number of both
unconstrained paths and paths satisfying the weight-product constraint
grows rapidly with path length $\ell$. This in turn causes running time
to increase, but at very different rates. At $\ell=2$ the two systems
are essentially tied (1.3$\times$), but the gap then widens to
13$\times$ at $\ell=3$ and 60$\times$ at $\ell=4$, where \MEMG already needs
over 5 minutes. At $\ell=5$, \MEMG exceeds the 30-minute timeout while
\ourabstraction completes in 869 sec.

Like in the experiments above, \ourabstraction's advantage comes from pushing
the weight-product constraint into path exploration, whereas \MEMG evaluates
it on fully assembled candidate paths, so its running time tracks the
unconstrained candidate count. Early filtering removes 58--77\% of the
length-$\ell$ candidates for $\ell=2,\ldots,4$; combined with set-at-a-time
execution, this yields speedups that grow with $\ell$.
\ourabstraction's own running time follows its output size. Growth is not
monotone: the per-length result count grows by $185\times$, $18.5\times$, and
$104\times$ from $\ell=2$ to $5$. This is a property of the graph rather than of
the bound: 69\% of edges have weight exactly $1.0$ and thus never tighten the
product, and paths that return to the high-degree start vertex fan out again
over its ${\sim}1{,}500$ admissible edges. The $161\times$ jump in running time from $\ell=4$ to $\ell=5$ is mostly explained by the $100\times$ growth in result size.

\section{Related Work}\label{sec:related}

Path queries over graphs have been extensively studied, with Regular Path Queries
(RPQs), which match edge-label sequences against regular
expressions~\cite{mendelzon1987graphical, barcelo2013querying}, as the foundational
pattern. The dominant evaluation paradigm is the product-graph
construction~\cite{Mendelzon}, where the input graph is traversed jointly with an
automaton representing the query, underlying a wide range of systems and
techniques~\cite{TRE_RPQ, KoschRPQsBG, arroyuelo2022time, belyanin2024single,
pacaci2020regular, bucchi2021core, ReutterVrgocRDFOverWeb, tetzel2019graph,
Wadhwa2019RSPQ, MDBVrgoc}. Some work extends RPQs with data (property) constraints
such as comparisons and aggregates~\cite{figueira2022data, libkin2016querying,
2012LibDomRPQSData}, primarily through richer automata models---register
automata~\cite{kaminski1994finite}, parametric automata~\cite{li2025answering},
variable automata~\cite{vrgovc2015using}, and Parikh
automata~\cite{figueira2015path}---centered on the tradeoff between expressiveness
and complexity. \ourabstraction's goal is different: it offers no
expressiveness-complexity characterization of the constraint language
\cite{losemann2013complexity}, because it
targets efficient execution of arbitrary constraints rather than delimiting a
tractable class.

Recent work on graph query languages and standards highlights the growing
importance of combining label patterns with property
constraints~\cite{libkin2025querying}. Yet, due to limited expressiveness of
pattern-matching clauses~\cite{gheerbrant2024gql}, these systems evaluate
``difficult'' property constraints at the end over materialized paths, often via
list-based constructs~\cite{gheerbrant2025dangers}. SOTA graph and relational DBMS
thus cannot identify early-filtering opportunities and produce large intermediate
results. This is the gap \ourabstraction closes: by exposing negatively
stable path constraints, it enables early filtering during traversal rather than
after full materialization---for any combination of label and property
constraints, which no prior approach supports generally.

A complementary line studies path enumeration under modes such as shortest, simple,
or trail paths~\cite{farias2023pathfinder, martensRepresentPath23}.
Other work explores strategies beyond left-deep query plans and
standard BFS-style path exploration~\cite{KoschRPQsBG, nguyen2020accelerating, nguyen2021efficiently, TRE_RPQ, yakovets2015waveguide, yakovets2016query}.
These are orthogonal to our work as \ourabstraction is designed to be
\emph{compatible}
with advanced approaches such as splitting of path patterns and performing
bi-directional search. In future work it would be interesting to
explore integrating \ourabstraction into approaches like
WAVEGUIDE by designing cost- and selectivity-estimation techniques
for it.

Relational approaches leverage recursive queries and joins for graph
querying. After early formalizations of RPQs via relational algebra with transitive
closure~\cite{yakovets2013evaluation, marton2017formalising},
recent work shows relational engines can outperform specialized graph DBMS on RPQs ~\cite{chlyah2025distributed,
relational_algebra_fejza2023mu, wolde2023duckpgq}, and systems such as
GRFusion~\cite{hassan_extending_2018}, GQ-Fast~\cite{lin_fast_2016},
GraphflowDB~\cite{jin2021making}, and DuckPGQ embed graph abstractions into
relational engines. Some support limited predicate pushdown but fail on
list-based constraints like those in \Cref{ex:q1_bitcoin,ex:full_recap}. \ourabstraction supports all these constraints and compiles them into recursive
relational queries with effective early filtering.

A recent line of research applies LLMs to automatically rewrite SQL
queries to improve their performance. Rule-selection
systems such as R-Bot \cite{rbot} and LLM-R\textsuperscript{2} \cite{llmr2}
use an LLM to choose and order rules from a given catalog (e.g., Calcite's),
so they cannot introduce a rewrite the catalog lacks---and no existing
catalog contains rules needed for our target queries. Systems
such as GenRewrite~\cite{genrewrite}, LITHE~\cite{lithe} and
QUITE~\cite{quite} can produce new rewrite rules, attempting to
demonstrate their correctness via various approaches including
iterative correction or result-equivalence testing over sampled databases.
In the end, only a formal proof will convince a user who cares about the
query result. As we discuss in \Cref{subsec:LLMCorrectness},
existing SQL equivalence provers~\cite{zhou2022spes, wang2022wetune, ding2023proving,tang2025proving} fall short for recursive queries.
A rewrite---by hand or LLM---of a recursive, property-constrained path query
therefore leaves a global equivalence obligation no current tool addresses.
With \ourabstraction this problem is reduced to fairly mechanical
inductive proofs of invariants on transition sequences in an NFA.

\section{Conclusions and Future Work}
\label{sec:conclusions}

We demonstrated that SOTA graph and relational DBMS struggle to push
list-based property
constraints deep into path-query processing. To address this issue, we
introduced the notion of negatively stable constraints and proposed
\ourabstraction as a clean abstraction for specifying the corresponding
early filtering opportunities. This reduced intermediate result
cardinality by a factor of up to 1 million and running time by a factor
of up to 230,000, even for simple queries and small graphs.
When a query does not admit early filtering, or when one cannot
identify negatively stable constraints,
\ourabstraction behaves like the SOTA competitors, thus providing
``regret-free'' optimization.
While LLMs can often produce improved SQL queries with minimal
prompting, a major challenge is to prove the equivalence to the original
query. With \ourabstraction this simplifies to fairly mechanical
induction proofs of state invariants along transition sequences
in a dictionary-augmented NFA.

By offering a clean abstraction for declaring functionality to merge
state across arbitrary path segments, \ourabstraction is naturally
compatible with advanced query plans that split path patterns and apply
bi-directional search. Exciting future work includes extending
the optimizers of SOTA relational DBMS so that they can support
such plans for recursive join queries. We are also interested
in extending our \ourabstraction workbench implementation
to directly integrate LLMs for automatically generating
the required functions and predicates for a given
path query.

\begin{acks}
This material is based upon work supported by the U.S. National Science Foundation Graduate Research Fellowship Program under Grant No. DGE-2439018. Any opinions, findings, and conclusions or recommendations expressed in this material are those of the author(s) and do not necessarily reflect the views of the National Science Foundation.
This work was done in part while Mirek Riedewald was a long-term visitor
at the Simons Institute for the Theory of Computing.
\end{acks}


\balance
\bibliographystyle{ACM-Reference-Format}
\bibliography{recap}



\end{document}